\documentclass[aip,reprint,pof]{revtex4-1}

\usepackage[utf8]{inputenc}
\usepackage[english]{babel}
\usepackage{graphics}
\usepackage{graphicx}
\usepackage{amsmath,amssymb,esint}
\usepackage{multirow}
\usepackage{subfig}
\usepackage{natbib}
\usepackage[usenames,x11names]{xcolor}
\usepackage[colorlinks,linkcolor=blue,citecolor=blue]{hyperref}
\usepackage{upgreek}
\usepackage{algorithm}
\usepackage{algpseudocode}

\usepackage{tikz}
\usetikzlibrary{shapes.geometric, arrows, intersections, through}
\usetikzlibrary{arrows.meta}
\usetikzlibrary{calc}
\tikzset{math3d/.style={z={(-0.65cm,-0.30cm)},y={(0cm,1cm)},x={(0.9cm,-0.15cm)}}}
\usetikzlibrary{shapes.geometric, arrows, intersections, through}
\usetikzlibrary{decorations.text, decorations.shapes, backgrounds}
\usetikzlibrary{arrows.meta}
\usetikzlibrary{colorbrewer}
\usetikzlibrary{patterns}

\renewcommand{\selectlanguage}[1]{}

\newcommand{\refereeone}[1]{{\color{black}{#1}}}
\newcommand{\refereefour}[1]{{\color{black}{#1}}}

\newcommand{\others}[1]{{\color{black}{#1}}}

\begin{document}


\title{Modeling \refereefour{time-delayed} acoustic interactions of cavitation bubbles and\\ bubble clusters}

\author{Pierre Coulombel}
\affiliation{\mbox{Department of Mechanical Engineering, Polytechnique Montr\'eal, Montr\'eal, H3T 1J4, Québec, Canada}}

\author{Fabian Denner}
\email[]{fabian.denner@polymtl.ca}
\affiliation{\mbox{Department of Mechanical Engineering, Polytechnique Montr\'eal, Montr\'eal, H3T 1J4, Québec, Canada}}


\begin{abstract}

We propose a low-dimensional modeling approach to simulate the dynamics, acoustic emissions and interactions of cavitation bubbles, based on a quasi-acoustic assumption. This quasi-acoustic assumption accounts for the compressibility of the medium surrounding the bubble and its finite speed of sound, whereby the potential of the acoustic wave emitted by the bubble propagates along outgoing characteristics. With these ingredients, a consistent set of equations describing the radial bubble dynamics as well as the resulting acoustic emissions and bubble-bubble interactions is obtained, which is accurate to the first order of the Mach number. This model is tested by considering several representative test cases, including the resonance behavior of multiple interacting bubbles and the response of dense mono- and polydisperse bubble clusters to a change in ambient pressure. The results are shown to be in excellent agreement with results reported in the literature. The differences associated with the finite propagation speed of the acoustic waves are observed to be most pronounced for the pressure-driven bubble dynamics in dense bubble clusters and the onset of cavitation in response to a change in ambient pressure.\\[0.5em]
\textsl{\small Copyright (2024) Author(s). This article is distributed under a Creative Commons Attribution (CC BY) License.}
\end{abstract}


\maketitle 

\section{Introduction}
Cavitation refers to the dynamics of pressure-driven bubbles, a complex phenomenon involved in a large number of engineering applications ranging from ultrasound imaging in medicine \cite{Matsumoto2005, Song2019} to the design of ship propellers \cite{Franc2005, Hsiao2015}. Understanding the physics of cavitation is 
essential, as the collapse of cavitation bubbles can focus a large amount of energy and, consequently, may cause serious damage to surrounding objects  \cite{Dular2019}, such as tissue \cite{Salzar2017}, cells in bioreactors \cite{Walls2017}, or solid walls \citep{Johnsen2009}. In most of these applications, bubbles are aggregated as clusters, whereby the bubbles interact with their neighbors through the generated acoustic emissions, leading to complicated dynamics that a single isolated bubble does not exhibit. For instance, the acoustic interactions delay the onset of cavitation by 
changing the pressure experienced by a bubble \cite{Ida2009} or promote a more violent bubble collapse than in the case of a single bubble \cite{Wang1999, Matsumoto2005, Shen2021, Deng2024}.

The radial dynamics of cavitation bubbles can be described by Rayleigh-Plesset (RP) equations, which are based on the seminal work of Lord Rayleigh on the collapse of an empty cavity in an incompressible liquid \citep{Rayleigh1917}.
These RP-type equations have since been extended to include the compressibility of the liquid \cite{Gilmore1952, Trilling1952,Keller1980, Prosperetti1986, Lezzi1987,Denner2021} as well as the interactions between bubbles due to their acoustic emissions \cite{Fuster2014, Fuster2019}, although RP-type equations typically assume that the bubbles remain spherical, which is a significant limitation in practice. Nevertheless, RP-type equations have been used successfully to study pressure-driven bubble dynamics \citep{Plesset1977,Lauterborn2010,Denner2024a}, including confinement \citep{Fu2023},  heat transfer \citep{Stricker2011,Zhou2020}, non-spherical behavior \citep{Klapcsik2019}, and other phenomena. Even though fully resolved two-/three-dimensional numerical simulations are frequently used today to study the pressure-driven behavior of individual and multiple bubbles \cite{Folden2023,Mnich2024, Saini2024}, RP-type equations are widely used to gain a more detailed understanding of the complex nonlinear pressure-driven dynamics of individual and multiple interacting bubbles \cite{Fuster2019, Folden2023}. Compared to fully resolved simulations, RP-type equations offer accurate solutions for spherical bubbles at a computational cost that is orders of magnitude smaller.

Using RP-type equations, \refereefour{the bubble-bubble interactions are typically modeled as instantaneous, even when using models that account for the compressibility of the surrounding medium, such as the
for instance 
Keller-Miksis \citep{Keller1980} or Gilmore \citep{Gilmore1952} models,
meaning that compressibility is not consistently 
taken into account.}
The RP-type equations governing the behavior of the bubbles, including the additional driving pressure induced by the emissions of the neighbor bubbles, can then be solved conveniently in a coupled nonlinear system of ordinary differential equations, which is not possible if  the finite propagation time between bubbles is to be considered. 
\refereefour{\citet{Fan2021} proposed a model where time-delayed interactions were 
considered as part of
an RP-type model,
which, however, is based on a linearized equation of motion and 
solved in the frequency domain.
\citet{Haghi2022} proposed to account for the time-delay associated with a finite speed of sound by treating the Keller-Miksis equation including an acoustic interaction term as a delay differential equation; the acoustic interactions are, however,  derived by considering the liquid to be incompressible.}
The simplifying assumption of an infinite propagation speed of acoustic emission is, in the context of bubble interactions, often justified by the short propagation time in comparison to the excitation period \citep{Mettin1997}. This approach has, for instance, been used to study the change in collective bubble response \citep{Shen2021} and emitted noise \citep{Jiang2012} compared to single-bubble systems, the nonlinear behavior of bubble clusters \citep{Jiang2017, Haghi2019}, the subharmonic threshold \citep{Guedra2017} and the destruction of encapsulated microbubbles \citep{Yasui2009} used in medical imaging.
In general, the interactions between bubbles in a bubble cluster hinder their expansion, resulting in bubble oscillations that are slower and reach smaller amplitudes \citep{DAgostino1989, Arora2007, Nasibullaeva2013, Zhao2020c}. This observation is explained by the fact that bubbles are acting as a low-pass filter \citep{Ohl2024}, resulting in a shielding effect with the bubbles located at the edge of a cluster affecting the strength of the incident pressure wave experienced by distal bubbles \citep{Wang1999, Bremond2006, Deng2024}. The presence of neighbor bubbles also influences the response of the bubbles of a polydisperse bubble cluster to a sudden drop in ambient pressure, with a lower pressure required for the smaller bubbles to grow explosively \citep{Ida2009}.  Moreover, a bubble cluster has several resonance frequencies, the largest being the one of a single bubble, also known as the Minneart frequency \citep{Minnaert1933}, as suggested in the review of \citet{Fuster2019}. \citet{Haghi2019} showed that in polydisperse bubble clusters, the larger bubbles dominate the oscillations of the smaller bubbles, forcing them to oscillate at the same frequency. 

In this article, we propose a consistent model of the radial dynamics, acoustic emissions, and acoustic interactions of pressure-driven bubble dynamics based on a quasi-acoustic assumption. This model assumes that the bubbles are spherical \refereeone{and 
stationary,} and the liquid surrounding the bubbles is compressible, and uses a Lagrangian wave tracking algorithms to model the acoustic emissions and interactions of the bubbles. Considering a variety of representative test cases, we demonstrate the capabilities of the proposed model for predicting the complex bubble-bubble interactions and resulting dynamic behavior in mono- and polydisperse bubble clusters.

This article is organized as follows. The governing equations are presented in Section \ref{sec:Equations} and the RP-type equation that governs the bubble dynamics is derived in Section \ref{sec:rp}. Subsequently, the models for the acoustic emissions and interactions are presented in Sections \ref{sec:emissions} and \ref{sec:Acoustic_interactions}, respectively. Section \ref{sec:results} presents a comprehensive validation of the proposed model and conclusions are drawn in Section \ref{sec:conclusions}. 

\section{Governing equations}
\label{sec:Equations}

\refereeone{We consider the bubbles to remain 
stationary
in space and only their radial dynamics 
are
considered.}
In spherical symmetry, the conservation of mass and momentum are given as
\begin{eqnarray}
    \frac{\partial \rho}{\partial t} + \frac{1}{r^2} \, \frac{\partial}{\partial r} \left(r^2 \rho u \right) = 0 \label{eq:continuity_original} \\
    \frac{\partial u}{\partial t} + u \frac{\partial u}{\partial r} = - \frac{1}{\rho} \frac{\partial p}{\partial r}, \label{eq:momentum}
\end{eqnarray}
where $t$ denotes time, $r$ is the radial coordinate, $u$ is the velocity, $p$ is the pressure and $\rho$ stands for the density of the liquid. 
Assuming that the liquid is isentropic, with $\text{d}\rho = \text{d}p/c^2$, where $c$ is the speed of sound of the liquid, and imposing a spatially invariant ambient pressure $p_\infty(t)$, the continuity equation, Eq.~\eqref{eq:continuity_original}, can be formulated as
\begin{equation}
    \frac{1}{c^2} \left( \frac{\partial p}{\partial t} - \frac{\partial p_\infty}{\partial t} + u \frac{\partial p}{\partial r}\right) + \frac{\rho}{r^2} \, \frac{\partial}{\partial r} \left(r^2 u \right) = 0. \label{eq:continuity}
\end{equation}
Following the work of \citet{Gilmore1952}, we invoke the {\it quasi-acoustic} assumption for the derivation of a set of equations governing the radial dynamics of a spherically symmetric bubble and the corresponding acoustic emissions in a compressible liquid.
The quasi-acoustic assumption stipulates that the Mach number $u/c$ is small, $(u/c)^2 \ll 1$, and that the density and speed of sound vary little and can be considered as constant.

Since the considered one-dimensional flow is irrotational, the velocity $u$ can be expressed by the potential $\psi$, 
\begin{equation}
    u = - \frac{\partial \psi}{\partial r}
    . \label{eq:velocitypotential}
\end{equation}
Contrary to classical acoustics, the flow velocity $u$ is not assumed to be negligible and, with the velocity expressed by its potential $\psi$, the conservation of momentum is expressed, after integrating Eq.~\eqref{eq:momentum} from $r$ to $\infty$, by the transient Bernoulli equation as
\begin{equation}
    -\frac{\partial \psi}{\partial t} + \frac{u^2}{2} = - \, \frac{p-p_\infty}{\rho},
    \label{eq:sphericalmomentumintegrated}
\end{equation}
which implies $\psi=0$ for $r \rightarrow \infty$. Inserting the expressions following from Eqs.~\eqref{eq:momentum} and \eqref{eq:sphericalmomentumintegrated} for $\partial p/\partial r$ and $p-p_\infty$, respectively, into the continuity Eq.~\eqref{eq:continuity}, yields the wave equation \citep{Denner2024a}
\begin{equation}
    \frac{\partial^2 \psi}{\partial t^2} - c^2 \, \nabla^2 \psi = \frac{\partial u^2}{\partial t} + \frac{u}{2} \frac{\partial u^2}{\partial r}. \label{eq:wave_1d}
\end{equation}
The nonlinear terms of order $\mathcal{O}(u^2)$ on the right-hand side are negligible under the quasi-acoustic assumption, since $u^2\ll c^2$.

The velocity potential in spherical symmetry is defined as $\phi=r \psi$. Assuming $u^2\ll c^2$, the wave equation of the velocity potential $\phi$ in spherical symmetry follows from Eq.~\eqref{eq:wave_1d} as
\begin{equation}
    \frac{\partial^2 \phi}{\partial t^2} - c^2 \, \frac{\partial^2 \phi}{\partial r^2} = 0,
    \label{eq:wave_spherical}
\end{equation}
and rearranging Eq.~\eqref{eq:sphericalmomentumintegrated} we obtain \citep{Gilmore1952}
\begin{equation}
g = \frac{\partial \phi}{\partial t} = r \left(\frac{p-p_\infty}{\rho}+\frac{u^2}{2} \right). \label{eq:g}
\end{equation}
By virtue of Eq.~\eqref{eq:wave_spherical}, both the potential $\phi$ and its temporal derivative $g = {\partial \phi}/{\partial t}$ have a constant amplitude and propagate along outgoing characteristics with speed $c$, such that they are described by the advection equation
\begin{equation}
    \frac{\partial \chi}{\partial t} + c \frac{\partial \chi}{\partial r} = 0,
    \label{eq:adv_phi-g}
\end{equation}
where $\chi \in \{ \phi, g\}$.

\section{Bubble dynamics}
\label{sec:rp}

An equation governing the radial dynamics of a spherical bubble in a compressible liquid is derived based on the quasi-acoustic assumption. Expanding Eq.~\eqref{eq:adv_phi-g} for $\chi=g$, we obtain
\begin{multline}
    \frac{r}{\rho} \left(\frac{\partial p}{\partial t} -\frac{\partial p_\infty}{\partial t} \right) + r u \frac{\partial u}{\partial t} + \frac{c r}{\rho} \frac{\partial p}{\partial r} + c r u \frac{\partial u}{\partial r} \\ + c \left( \frac{p-p_\infty}{\rho} + \frac{u^2}{2}\right) = 0. \label{eq:advection3}
 \end{multline}
From the conservation of momentum, Eq.~\eqref{eq:momentum}, and the conservation of mass, Eq.~\eqref{eq:continuity}, 
the spatial derivatives of the flow velocity and the pressure at the bubble wall follow as \citep{Trilling1952,Denner2024a}
\begin{align}
    \left. \frac{\partial u}{\partial r} \right|_{r=R} &= - \frac{\dot{p}_\text{L}-\dot{p}_\infty}{\rho c^2} - 2 \frac{\dot{R}}{R}\label{eq:u_r}\\
    \left. \frac{\partial p}{\partial r} \right|_{r=R} &= - \rho \ddot{R} \label{eq:p_r}.
\end{align}
Evaluating Eq.~\eqref{eq:advection3} at the bubble wall ($r=R$), with the spatial derivatives given in Eqs.~\eqref{eq:u_r} and \eqref{eq:p_r}, reads as
\begin{multline}
    R \left( 1 - \frac{\dot{R}}{c} + \frac{\dot{R}^2}{c^2} \right) \frac{\dot{p}_\text{L}-\dot{p}_\infty}{\rho} +  \left(2 \dot{R} - c \right) R \ddot{R} \\ + \frac{3}{2} \left(\frac{4 \dot{R}}{3} - c \right) \dot{R}^2 + c \, \frac{p_\text{L}-p_\infty}{\rho} = 0. \label{eq:bubblegov1}
\end{multline}
Multiplying Eq.~\eqref{eq:bubblegov1} by $(1+\dot{R}/c)/c$, as suggested by \citet{Gilmore1952}, yields 
\begin{multline}
    \left(1- \frac{\dot{R}}{c} - 2 \frac{\dot{R}^2}{c^2} \right) R \ddot{R}  + \frac{3}{2} \left(1 - \frac{\dot{R}}{3c} - \frac{4 \dot{R}^2}{3 c^2} \right) \dot{R}^2 \\ =  \left(1 + \frac{\dot{R}}{c} \right) \frac{p_\text{L}-p_\infty}{\rho} +   \left(1 + \frac{\dot{R}^3}{c^3}\right) \frac{R}{c}  \, \frac{\dot{p}_\text{L}-\dot{p}_\infty}{\rho} . \label{eq:bubblegov2}
 \end{multline}
Neglecting all high-order terms of the Mach number $(\dot{R}/c)^\alpha$ with $\alpha \geq 2$, which is justified by $(\dot{R}/c)^2 \ll 1$ inherent to the underpinning quasi-acoustic assumption, the governing equation for the bubble radius $R$ follows as
\begin{multline}
    \left(1- \frac{\dot{R}}{c} \right) R \ddot{R}  + \frac{3}{2} \left(1 - \frac{\dot{R}}{3c} \right) \dot{R}^2 \\ =  \left(1 + \frac{\dot{R}}{c} \right) \frac{p_\text{L}-p_\infty}{\rho} + \frac{R}{c}\frac{\dot{p}_\text{L}-\dot{p}_\infty}{\rho}. \label{eq:km}
\end{multline}
This equation represents the widely-used Keller-Miksis equation \citep{Keller1980} and is accurate to first order in the Mach number $\dot{R}/c$ of the bubble wall\citep{Prosperetti1984a}.

The liquid pressure at the bubble wall incorporates the kinematic boundary conditions at the bubble wall, and is defined as
\begin{equation}
    p_\text{L} = p_\text{G} - \frac{2 \sigma}{R} - 4 \, \mu \frac{\dot{R}}{R}, \label{eq:pL}
\end{equation}
where  $\sigma$ is the surface tension coefficient of the gas-liquid interface and $\mu$ is the dynamic viscosity of the liquid. The gas pressure inside the bubble, $p_\text{G}$, assuming heat and mass transfer between the gas and the liquid are negligible, is defined as 
\begin{equation}
p_\text{G} = p_\text{G,0} \left(\frac{R_0^3}{R^3} \right)^{\kappa}, \label{eq:pG}
\end{equation}
where $R_0$ is the initial bubble radius, $p_\text{G,0}$ is the gas pressure at $R_0$ and $\kappa$ is the polytropic exponent of the gas.

\section{Acoustic emissions}
\label{sec:emissions}

To analyze the acoustic emissions of individual bubbles and bubble clusters, as well as to model the acoustic interaction between bubbles, a suitable description of the pressure waves emitted and flow field induced by bubble oscillations is required.
Inserting the velocity potential $\psi = \phi/r$ into Eq.~\eqref{eq:velocitypotential} yields
\begin{equation}
    u = \frac{\phi}{r^2} - \frac{1}{r} \frac{\partial \phi}{\partial r}
\end{equation}
and, after inserting Eq.~\eqref{eq:adv_phi-g} with $\chi = \phi$, the velocity is given by
\begin{equation}
    u = \frac{\phi}{r^2} + \frac{g}{r \, c}. \label{eq:u_general}
\end{equation}
The pressure is obtained by rearranging Eq.~\eqref{eq:sphericalmomentumintegrated} as
\begin{equation}
    p = p_\infty + \rho \left( \frac{g}{r} -  \dfrac{u^2}{2} \right). \label{eq:p_general}
\end{equation}
Given meaningful definitions of $\phi$ and $g$, Eqs.~\eqref{eq:u_general} and \eqref{eq:p_general} describe the spherically symmetric flow field produced by the radial motion of a spherical bubble \citep{Gilmore1952,Denner2023,Denner2024a} and are, like Eq.~\eqref{eq:km}, derived consistently under the quasi-acoustic assumption.

\begin{figure}
    \centering
    \includegraphics[width=\linewidth]{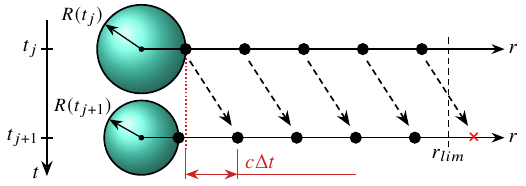}
    \caption{Schematic illustration of the Lagrangian transport of the acoustic emissions. The local radius $r(t)$, velocity $u(r,t)$ and pressure $p(r,t)$ represent together with the invariants $\phi(\tau)$ and $g(\tau)$ an {\it emission node} that is transported with the speed of sound $c$ and updated at each discrete time instance $t$. A node that has passed a predefined maximum radial coordinate, $r_\mathrm{lim}$, is considered obsolete and deleted.}
    \label{fig:lagrangianschematic}
\end{figure}

Following our previous work \citep{Denner2023}, the acoustic emissions of a cavitation bubble are tracked along outgoing characteristics with the speed of sound as \textit{emission nodes}, as illustrated in Figure \ref{fig:lagrangianschematic}. Each emission node contains information about the radial location $r(t)$, the pressure and velocity fields $p(r,t)$ and $u(r,t)$, and the invariants $\phi$ and $g$ of the acoustic emissions that follow from the employed model for the bubble dynamics. The radial coordinate $r(t)$ of an emission node is updated at each time $t$ by 
\begin{equation}
r(t)= R(\tau) + c \int_{\tau}^t  \text{d}t \approx R(\tau) +  c \sum_{i=1}^T \Delta t_i,
\end{equation}
where $T$ is the number of time steps $\Delta t$ from $\tau$ to $t$, and  $\tau$ is the retarded time $\tau=t-r/c$, representing the time at which the emission node is emitted at the bubble wall.
The invariants $\phi$ and $g$ at the time of emission $\tau$ are readily obtained by evaluating Eq.~\eqref{eq:u_general} and Eq.~\eqref{eq:p_general} at $r(t)=R(\tau)$ as
\begin{align}
\phi(\tau) &= R(\tau)^2 \dot{R}(\tau) - R(\tau) \,\frac{g(\tau)}{c} \label{eq:phi_qa}\\
g(\tau) &= R(\tau) \left[ \frac{p_\mathrm{L}(\tau)-p_\infty(\tau)}{\rho} + \frac{\dot{R}(\tau)^2}{2}\right]. \label{eq:g_qa}
\end{align}
Inserting Eqs.~\eqref{eq:phi_qa} and \eqref{eq:g_qa} into Eqs.~\eqref{eq:u_general} and \eqref{eq:p_general} then yields closed-form expressions for the local flow velocity and pressure \citep{Denner2023} that satisfy the boundary condition at the bubble wall by reducing to ${u}(R,\tau) = \dot{R}(\tau)$ and  ${p}(R,\tau) = p_{\text{L}} (\tau)$, respectively. 

\section{Acoustic interactions}
\label{sec:Acoustic_interactions}

When the interaction of multiple bubbles is to be considered, the velocity and pressure contributions of all bubbles in a suitably sized neighborhood ought to be considered. 
Assuming a superposition of the potential $\phi$ in the liquid \citep{Fuster2011a,Zhang2023b}, the local flow velocity and pressure at location $\mathbf{x}$ in three-dimensional space are given as 
\begin{align}
    u(\mathbf{x},t) &= \sum_{i=1}^{N} \frac{\phi_i(\tau)}{\Delta x_i^2} + \frac{1}{c} \sum_{i=1}^{N} \frac{g_i(\tau)}{\Delta x_i} \label{eq:u_inter} \\
    \Delta p(\mathbf{x},t) &=  \rho \left[ \sum_{i=1}^{N} \frac{g_i(\tau)}{\Delta x_i} - \frac{u(\mathbf{x},t)^2}{2} \right], \label{eq:dp_x} 
\end{align}
respectively, where $N$ is the number of bubbles and $\Delta x_i = |\mathbf{x} - \mathbf{x}_i|$ is the distance of the center $\mathbf{x}_i$ of bubble $i$ to the location $\mathbf{x}$. 

The driving pressure $p_\infty$ of each bubble is represented by the pressure in the liquid as if the considered bubble was absent, based on the ambient pressure $p_0$, the acoustic excitation pressure $ p_\text{a}$ (e.g., an ultrasound field), and the pressure contributed by the acoustic emissions of the neighbor bubbles. For bubble $i$, the driving pressure is, therefore, defined 
as 
\begin{multline}
    p_{{\infty},i}(t) = p_\infty(\mathbf{x}_i,t) = p_0 + p_{\text{a}}(\mathbf{x}_i,t) \\+ \underbrace{\rho \sum_{j=1}^{N_i} \frac{g_j(\tau)}{\Delta x_{ij}} - \frac{\rho}{2} \left[ \sum_{j=1}^{N_i} \frac{\phi_j(\tau)}{\Delta x_{ij}^2} + \frac{1}{c} \sum_{j=1}^{N_i} \frac{g_j(\tau)}{\Delta x_{ij}}\right]^2}_{p_\mathrm{inter}} , \label{eq:pinf_bubble}
\end{multline}
with $\Delta x_{ij} = |\mathbf{x}_i - \mathbf{x}_j|$ and where $p_\mathrm{inter}$ is the interaction pressure. 
The externally imposed acoustic excitation is either assumed to be spatially varying, representing a traveling wave, or spatially invariant, whereby all bubbles experience the same external acoustic excitation.
The number of considered neighbor bubbles, $N_i$, may include all bubbles of a population, as considered in this study, or only the bubbles in some predefined neighborhood of bubble $i$. Additional details about the numerical computation of the interaction terms in Eq.~\eqref{eq:pinf_bubble} are given in Appendix \ref{sec:computinginteractionterms}.

With the driving pressure defined by Eq.~\eqref{eq:pinf_bubble}, the complete Keller-Miksis equation solved for each bubble $i$ follows from Eq.~\eqref{eq:km} as
\begin{multline}
    \left(1- \frac{\dot{R}_i}{c} \right) R_i \ddot{R}_i  + \frac{3}{2} \left(1 - \frac{\dot{R}_i}{3c} \right) \dot{R}_i^2 \\ =  \left(1 + \frac{\dot{R}_i}{c} \right) \frac{p_{\text{L},i}- p_{{\infty},i}}{\rho} + \frac{R_i}{c} \frac{\dot{p}_{\text{L},i}-\dot{p}_{\infty,i}}{\rho}. \label{eq:km_bubble}
 \end{multline}

\section{Results}
\label{sec:results}

To validate the proposed model for the dynamics and acoustic interactions of cavitation bubbles and bubble clusters, and to highlight the differences to previous work, we consider the frequency response of polydisperse microbubbles (Sec.~\ref{sec:Haghi2019}), the resonance pattern of a bubble screen (Sec.~\ref{sec:Fan2021}), the asymmetric collapse of a spherical bubble cluster (Sec.~\ref{sec:Maeda2019}), as well as the pressure-induced onset of cavitation in monodisperse and polydisperse bubble clusters (Sec.~\ref{sec:Ida2009} and Sec.~\ref{sec:tension}).
The proposed methodology is implemented in the open-source software library {\tt APECSS} \citep{Denner2023a}, of which the version (v1.7) used to produce the presented results is available at \url{https://doi.org/10.5281/zenodo.13850831}.

\subsection{Frequency response of polydisperse microbubbles}
\label{sec:Haghi2019}

\begin{figure*}
    \centering
    \includegraphics[scale=0.95]{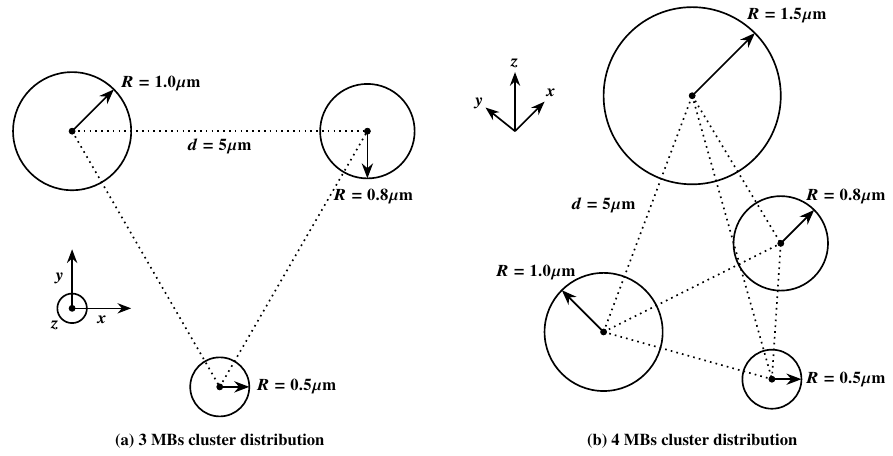}
    \caption{Schematic illustration of the two bubble clusters used for the frequency response analysis, \others{following the work of \citet{Haghi2019}}: (a) three bubbles located on the corners of a regular triangle and (b) four bubbles located on the corners of a regular tetrahedron. In both clusters, the bubbles are located $d = 5 \ \upmu\mathrm{m}$ apart from each other.}
    \label{fig:Haghi2019_Clusters}
\end{figure*}

\citet{Haghi2019} studied the resonance behavior of a small number interacting bubbles using a simplified Keller-Miksis equation, 
\begin{equation}
    \begin{split}
    \left(1- \frac{\dot{R}_i}{c} \right) R_i \ddot{R}_i  + \frac{3}{2} \left(1 - \frac{\dot{R}_i}{3c} \right) \dot{R}_i^2 = \frac{p_{\text{L},i}- p_0 - p_{\text{a}}(\mathbf{x}_i,t)}{\rho} \\ - \sum_{j=1}^{N} \left( \frac{R_{j}(t)^{2} \Ddot{R}_{j}(t) + 2 R_{j}(t) \Dot{R}_{j}(t)^{2}}{\Delta x_{ij}} - \frac{R_{j}(t)^{4} \Dot{R}_{j}(t)^{2}}{2 \Delta x_{ij}^{4}}\right),
    \label{eq:pinf_bubble_instant_haghi}
    \end{split}
 \end{equation}
where the second term on the right-hand side describes the bubble-bubble interactions. These interactions are assumed to be instantaneous, which is equivalent to assuming an incompressible liquid. They found that the dynamics of the smaller bubbles in a polydisperse cluster are strongly influenced by their neighbors of larger size, leading to two distinct scenarios:
\begin{enumerate}
    \item Constructive interactions, whereby the oscillations of smaller and larger bubbles are in phase, leading to amplified oscillations of the smaller bubbles. 
    \item Destructive interactions, whereby the resonance modes of the smaller bubbles are suppressed and the oscillation dynamics of the smaller bubbles are governed by the larger bubbles in their neighborhood.
\end{enumerate}

\begin{figure*}
    \centering
    \includegraphics[width=0.75\textwidth]{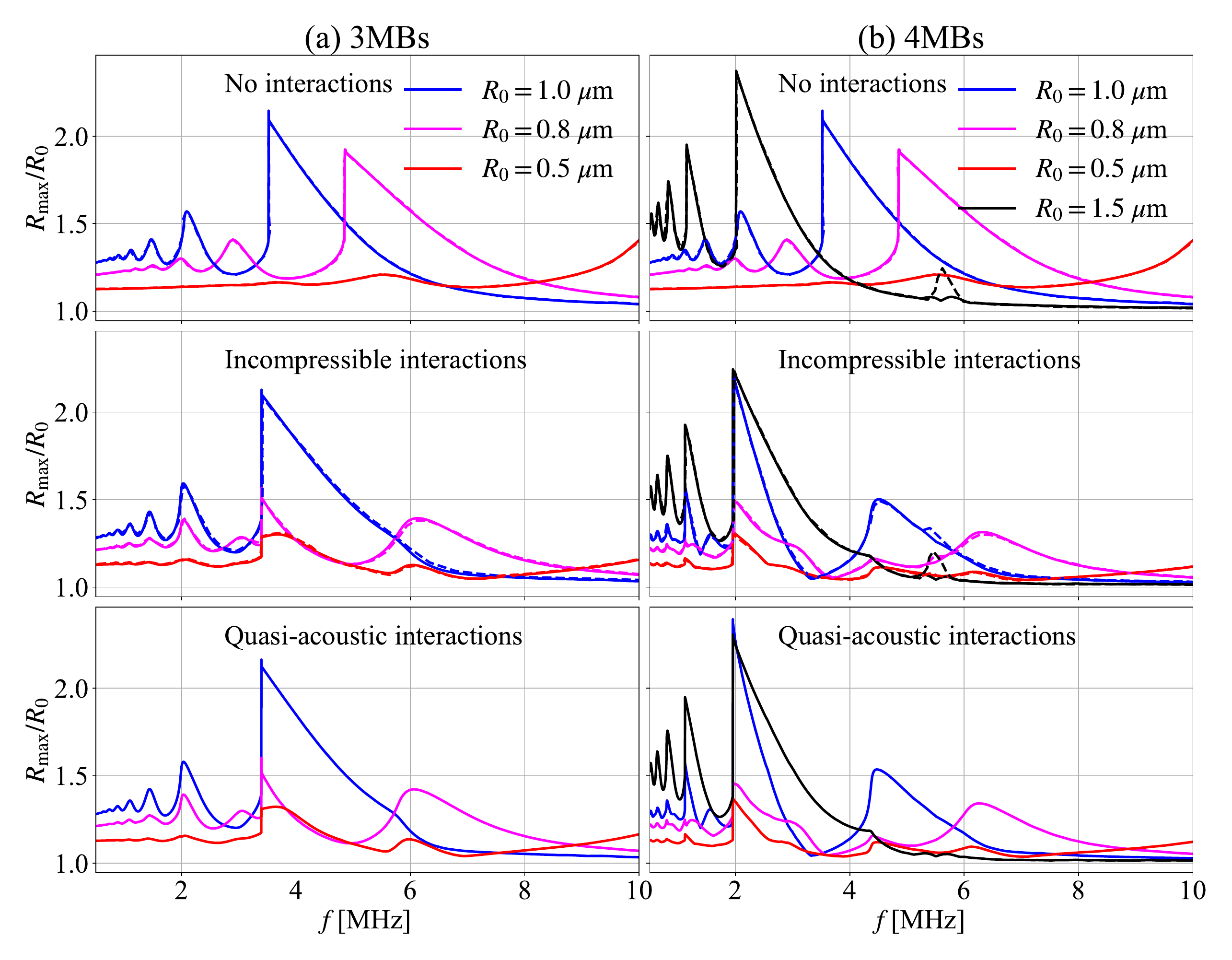}
    \caption{The normalized maximum radius attained by each bubble  
     as a function of the excitation frequency $f$ for (a) the three-bubble cluster and (b) the four-bubble cluster illustrated in Figure \ref{fig:Haghi2019_Clusters}, in response to a sinusoidal pressure excitation with an amplitude of $\Delta p = 120 \ \mathrm{Pa}$. 
    Dashed lines show the results reported by \citet{Haghi2019}.}
    \label{fig:Haghi2019_Results}
\end{figure*}

\begin{figure*}
    \centering
    \includegraphics[width=0.75\textwidth]{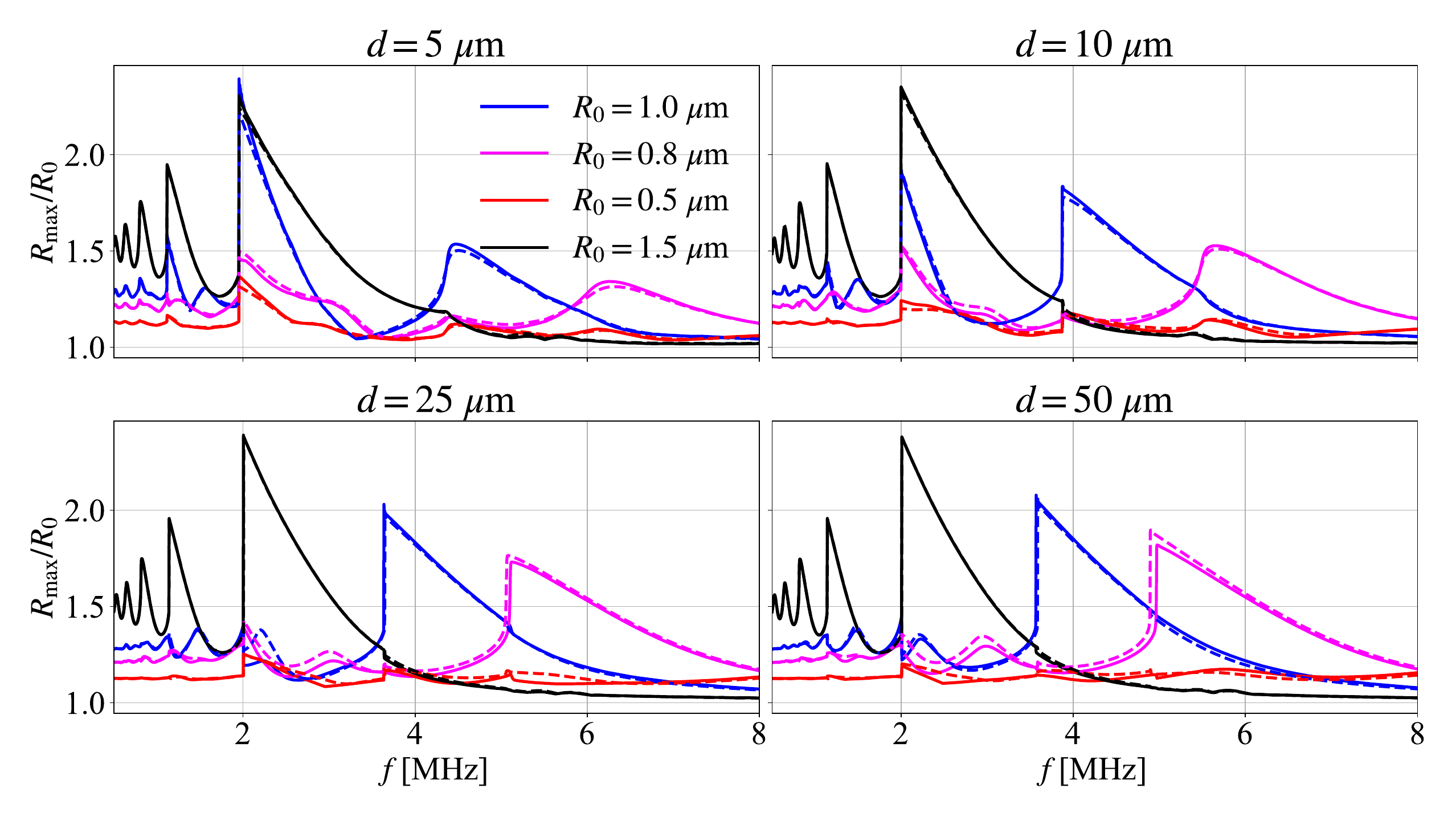}
    \caption{
    The normalized maximum radius attained by each bubble 
    as a function of the excitation frequency $f$ for the four-bubble cluster illustrated in Figure \ref{fig:Haghi2019_Clusters} with different bubble separation distances $d$, in response to a sinusoidal pressure excitation with an amplitude of $\Delta p = 120 \ \mathrm{Pa}$. Dashed lines represent incompressible interactions, while solid lines represent quasi-acoustic interactions.}
    \label{fig:Haghi2019_Distance}
\end{figure*}

Because we want to compare the proposed quasi-acoustic model with the incompressible approach, we will use a more complete Keller-Miksis equation for our computations with the incompressible interaction term, defined for bubble $i$ as
 \begin{multline}
    \left(1- \frac{\dot{R}_i}{c} \right) R_i \ddot{R}_i  + \frac{3}{2} \left(1 - \frac{\dot{R}_i}{3c} \right) \dot{R}_i^2 \\= \left(1 + \frac{\dot{R}_i}{c} \right) \frac{p_{\text{L},i} - p_{\infty,i,\mathrm{IC}}}{\rho} + \frac{R_i}{c} \frac{\dot{p}_{\text{L},i}-\dot{p}_{\text{a}}(\mathbf{x}_i,t)}{\rho}, \label{eq:pinf_bubble_instant}
 \end{multline}
 where 
 \begin{multline}
 p_{\infty,i,\mathrm{IC}} = p_{0} + p_{\text{a}}(\mathbf{x}_i,t) \\+ \rho \sum_{j=1}^{N} \left( \frac{R_{j}(t)^{2} \Ddot{R}_{j}(t) + 2 R_{j}(t) \Dot{R}_{j}(t)^{2}}{\Delta x_{ij}} - \frac{R_{j}(t)^{4} \Dot{R}_{j}(t)^{2}}{2 \Delta x_{ij}^{4}}\right).
 \end{multline}

Following the work of \citet{Haghi2019}, we consider the frequency response of two bubble clusters, schematically illustrated in Figure \ref{fig:Haghi2019_Clusters}. The first cluster is composed of three bubbles with initial radii of $1.0 \ \upmu\mathrm{m}$, $0.8 \ \upmu\mathrm{m}$, and $0.5 \ \upmu\mathrm{m}$, located $d = 5 \ \upmu\mathrm{m}$ apart from each other on the corners of a regular triangle. The second cluster includes an additional bubble with a radius of $1.5 \ \upmu\mathrm{m}$, with the four bubbles located on the corners of a regular tetrahedron. The bubbles are assumed to contain air, with polytropic exponent $\kappa = 1.4$, and are situated in water, with a density of $\rho = 1000 \ \mathrm{kg}/\mathrm{m}^{3}$, a speed of sound of $c = 1500 \ \mathrm{m}/\mathrm{s}$, a dynamic viscosity of $\mu = 1.002 \times 10^{-3} \ \mathrm{Pa}\, \mathrm{s}$, and a surface tension coefficient of $\sigma = 0.0728 \ \mathrm{N}/\mathrm{m}$. The ambient pressure is $p_{0} = 10^{5} \ \mathrm{Pa}$.
The bubble clusters are excited with a progressive sinusoidal wave defined as
\begin{equation}
    p_{\text{a}}(\mathbf{x},t) = \Delta p \sin \left( 2 \pi f \left[ t - \frac{x}{c} \right] \right) ,
\end{equation}
with a pressure amplitude of $\Delta p = 120 \ \mathrm{kPa}$, a frequency ranging from $0.5 \ \mathrm{MHz}$ to $10.0 \ \mathrm{MHz}$, and 
where $x$ represents the $x$-coordinate as illustrated in Figure \ref{fig:Haghi2019_Clusters}. For each frequency, the maximum radius achieved by each bubble in periodic steady state is recorded.

The maximum bubble radii as a function of the excitation frequency are shown in Figure \ref{fig:Haghi2019_Results}, without bubble-bubble interactions (i.e., each bubble is considered as an isolated bubble), with incompressible bubble-bubble interactions (i.e., assuming the liquid speed of sound is infinite), and with quasi-acoustic bubble-bubble interactions (i.e., considering a finite liquid speed of sound). The results without interactions and with incompressible interactions are in excellent agreement with the results of \citet{Haghi2019}. 
When acoustic interactions between the bubbles are considered, either incompressible or quasi-acoustic interactions, the largest bubble imposes its oscillation behavior onto the smaller bubbles, suppressing
the resonance response of the smaller bubbles.
However, the resulting differences between incompressible and quasi-acoustic interactions are small, which we attribute to the rather small separation distance of the bubbles of only $d = 5 \ \upmu\mathrm{m}$, leading to nearly instantaneous interactions even when the finite propagation time of the acoustic emissions is considered with the proposed quasi-acoustic model.

Figure \ref{fig:Haghi2019_Distance} highlights the influence of the bubble-bubble distance. 
Increasing the distance separating the bubbles leads to less pronounced interactions, with a gradual recovery of 
the results obtained without considering interactions shown in Figure \ref{fig:Haghi2019_Results}. An exception is the smaller bubble, which remains dominated by all its neighbors, not just the largest bubble. In addition, the differences in the bubble dynamics predicted with quasi-acoustic interactions compared to the incompressible interactions becomes more apparent when increasing the distance between the bubbles, especially for the smaller bubbles in the low-frequency range.

\subsection{Resonance patterns of a monodisperse bubble screen}
\label{sec:Fan2021}

\citet{Fan2021} studied the importance of accounting for the compressibility of the liquid and the associated finite propagation speed of acoustic waves for the bubble-bubble interaction. 
They demonstrated that the time delay of the bubble-bubble interactions in such clusters results in the formation of spatial patterns.

\begin{figure}
    \centering
    \includegraphics[scale=1.0]{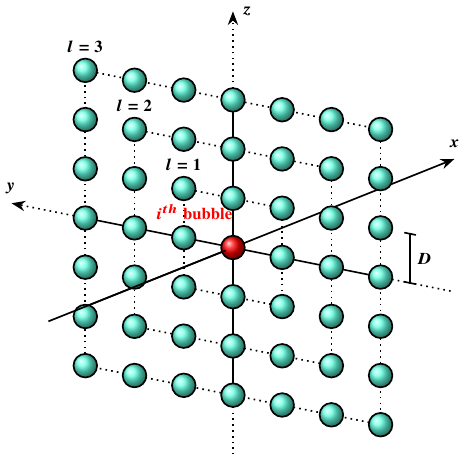}
    \caption{Schematic representation of the considered monodisperse bubble screen, following the work of \citet{Fan2021}.}
    \label{fig:Fan2021_BubblyScreen}
\end{figure}

Following the work of \citet{Fan2021}, we consider the  bubble screen shown in Figure \ref{fig:Fan2021_BubblyScreen}, consisting
of 51 $\times$ 51 uniformly spaced bubbles that are excited by
\begin{equation}
    p_{\text{a}}(t) = \Delta p \sin \left( \omega t\right),
\end{equation}
where $\omega$ is the angular frequency.
Each bubble has an initial radius of $R_{0}=1 \ \upmu\mathrm{m}$ and the bubbles are located at a distance of $D=400 R_{0}$. We consider an air-water system with a liquid density of $\rho = 1000 \ \mathrm{kg}/\mathrm{m}^{3}$, a speed of sound of $c = 1500 \ \mathrm{m}/\mathrm{s}$, an ambient pressure of $p_{0} = 10^{5} \ \mathrm{Pa}$, an initial gas pressure of $p_{\mathrm{G},0} = 10^{5} \ \mathrm{Pa}$, and a polytropic exponent of the gas of $\kappa = 1.4$. 
This choice of parameters leads to a Minneart resonance frequency \citep{Minnaert1933} for each individual bubble of $\omega_{0} = \sqrt{3 \kappa p_{0} / \left( \rho R_{0}^{2} \right)} = 20.49 \times 10^{6} \ \mathrm{rad/s}$. The bubble screen is excited with an angular frequency of $0.5 \leq \omega/\omega_{0} \leq 1.5$ and a pressure amplitude of $\Delta p = 100 \ \mathrm{Pa}$. 

\begin{figure*}
    \centering
    \includegraphics[width=\textwidth]{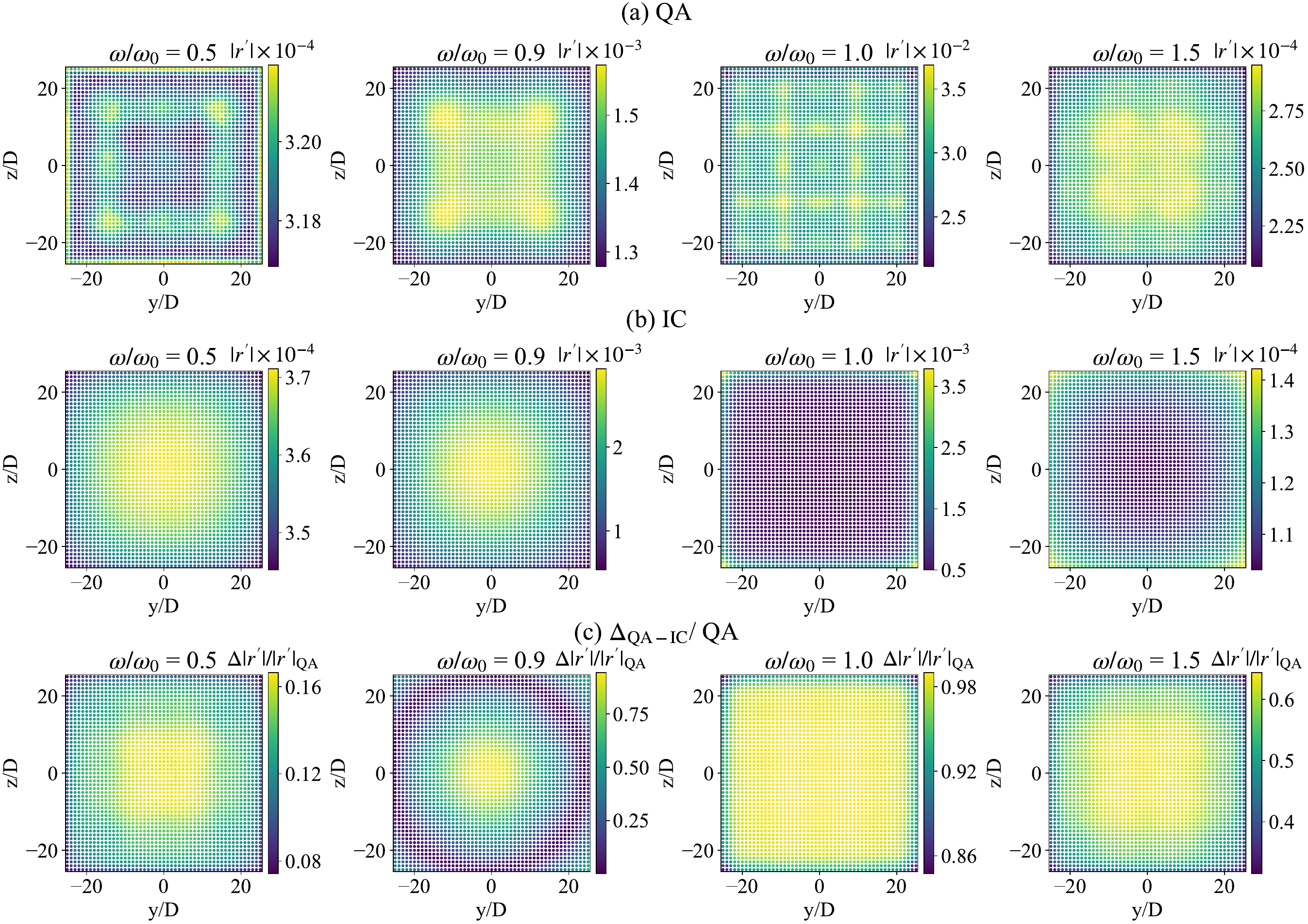}
    \caption{Dimensionless radial oscillation amplitude $|r'|$ at periodic steady state for each bubble in the monodisperse bubble screen, modeling the bubble-bubble interactions using (a) the proposed quasi-acoustic model (QA) and 
    (b) incompressible bubble-bubble interactions (IC),  \refereeone{and (c) shows the relative difference between the two models, with $\Delta |r'| = ||r'|_{\mathrm{QA}} - |r'|_{\mathrm{IC}}|$.} The bubble screen is excited by a sinusoidal wave with a pressure amplitude of $\Delta p = 100 \ \mathrm{Pa}$ and excitation frequency $\omega$. The interbubble distance is $D=400 R_{0}$, with $R_{0}=1.0 \ \upmu\mathrm{m}$ being the initial radius of the bubbles. The Minneart resonance frequency of the bubbles is $\omega_{0} = 20.49 \times 10^{6} \ \mathrm{rad/s}$.}
    \label{fig:Fan2021_Compressible}
\end{figure*}

\begin{figure*}
    \centering
    \includegraphics[width=0.7\linewidth]{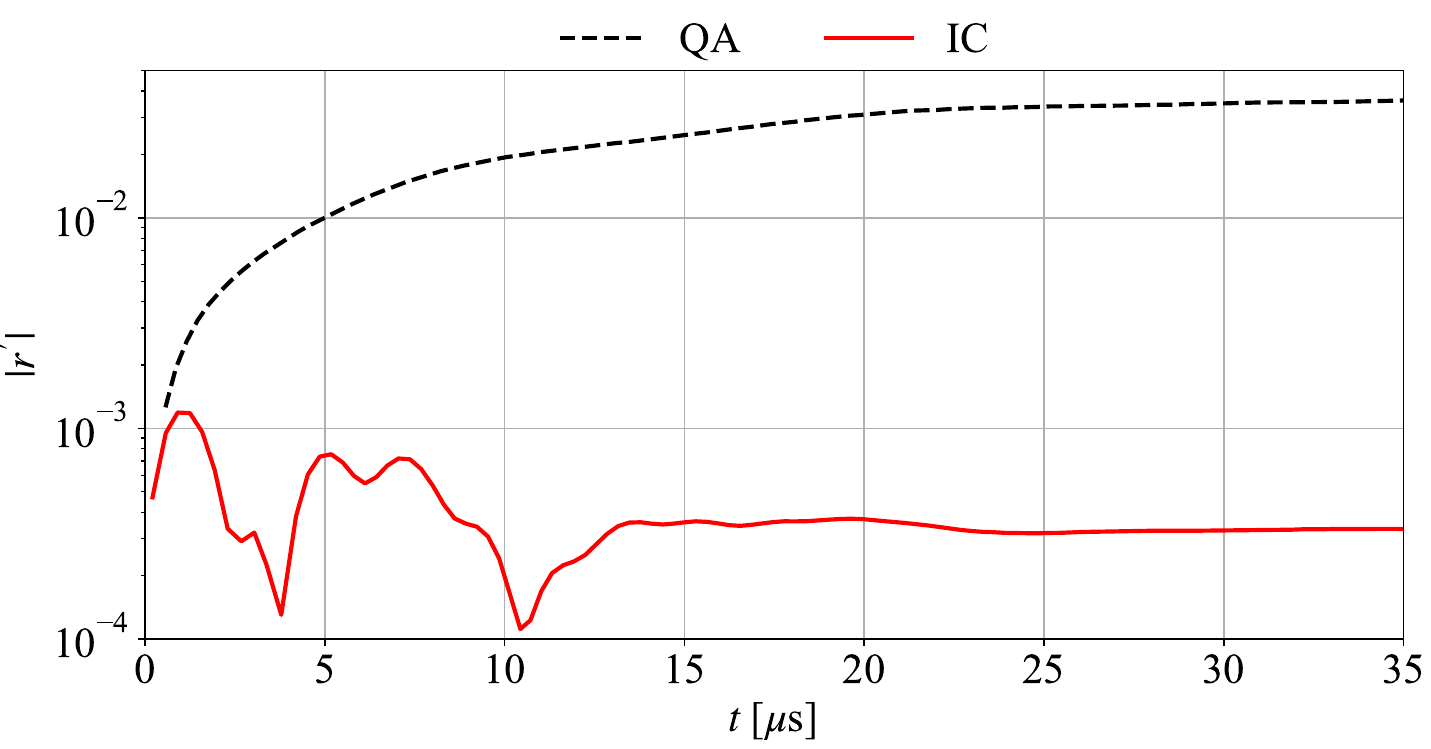}
    \caption{Evolution of the dimensionless radial oscillation amplitude $|r'|$  
    for the bubble at the center of the monodisperse bubble screen shown in Figure \ref{fig:Fan2021_Compressible}, excited with $\omega=\omega_0$.} 
    \label{fig:Fan2021_radius_center}
\end{figure*}

\begin{figure*}
    \centering
    \includegraphics[width=\textwidth]{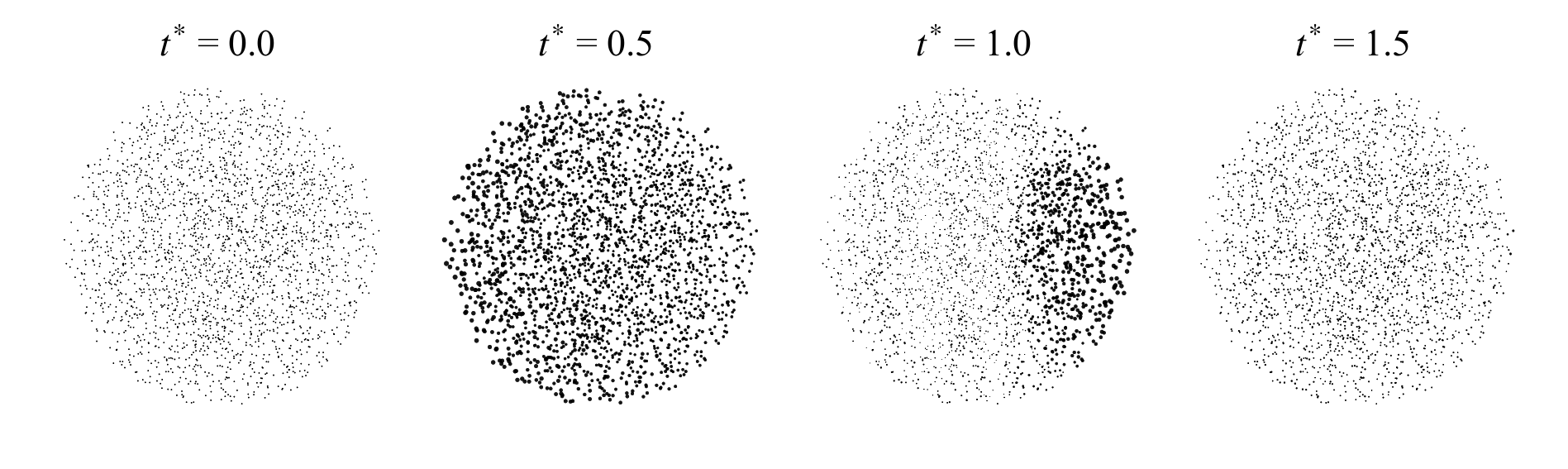}
    \caption{Evolution of a monodisperse spherical bubble cluster with radius $R_\mathrm{C} = 2.5 \ \mathrm{mm}$, composed of $N = 2500$ bubbles. The cluster is excited by a single pulse of a sinusoidal wave propagating from left to right, with a pressure amplitude of $\Delta p = -2.5 \ \mathrm{bar}$ and a frequency of $f = 50.0 \ \mathrm{kHz}$. The initial radius of each bubble is $R_{0} = 10 \ \upmu\mathrm{m}$ and
    $t^{*}=tf$ denotes the dimensionless time.}
    \label{fig:Maeda2019_ClusterEvolution}
\end{figure*}

The dimensionless radial oscillation amplitude $|r'| = \left( R_{\mathrm{max}} - R_{\mathrm{min}} \right)/2R_{0}$ of each bubble in the bubble screen at periodic steady state is shown in Figure \ref{fig:Fan2021_Compressible} for four excitation frequencies, with $R_{\mathrm{max}}$ and $R_{\mathrm{min}}$ representing respectively the maximum and minimum radius value during an oscillation period.
The results obtained with the quasi-acoustic model are in good agreement with the results reported by \citet{Fan2021},
especially with respect to the orders of magnitude of the oscillation amplitude. Qualitative differences in the amplitude pattern can be observed between our results and the results of \citet{Fan2021}, which we attribute to differences in the chosen solution domain.
While the proposed model solves the transient bubble dynamics and interactions in the time domain, \citet{Fan2021} solved the bubble dynamics and interactions in the frequency domain, thereby enforcing time periodicity. 

Considering instead incompressible bubble-bubble interactions, described by Eq.~\eqref{eq:pinf_bubble_instant},
the oscillations of the bubble screen show clear differences, see Figure \ref{fig:Fan2021_Compressible}, with respect to the magnitude of the oscillations and the amplitude patterns.  
\refereeone{For $\omega=\omega_{0}$, the relative difference between the two models $\Delta |r'| / |r'|_{\mathrm{QA}} = ||r'|_{\mathrm{QA}} - |r'|_{\mathrm{IC}}| / |r'|_{\mathrm{QA}} \approx 0.98$ for the majority of the bubbles. 
This
means that the radial oscillation amplitude for the incompressible model represents only 2 \% of the value obtained with the quasi-acoustic model. 
This difference of approximately two orders of magnitude is 
also visible 
in
Figure \ref{fig:Fan2021_radius_center} for the center bubble.}
In addition, the bubbles require a longer time to reach a periodic steady state when considering quasi-acoustic interactions ($\approx 30 \ \mu\mathrm{s}$) as compared to when considering incompressible interactions ($\approx 20 \ \mu\mathrm{s}$), see Figure \ref{fig:Fan2021_radius_center}.

\begin{figure*}
    \centering
    \includegraphics[width=0.7\textwidth]{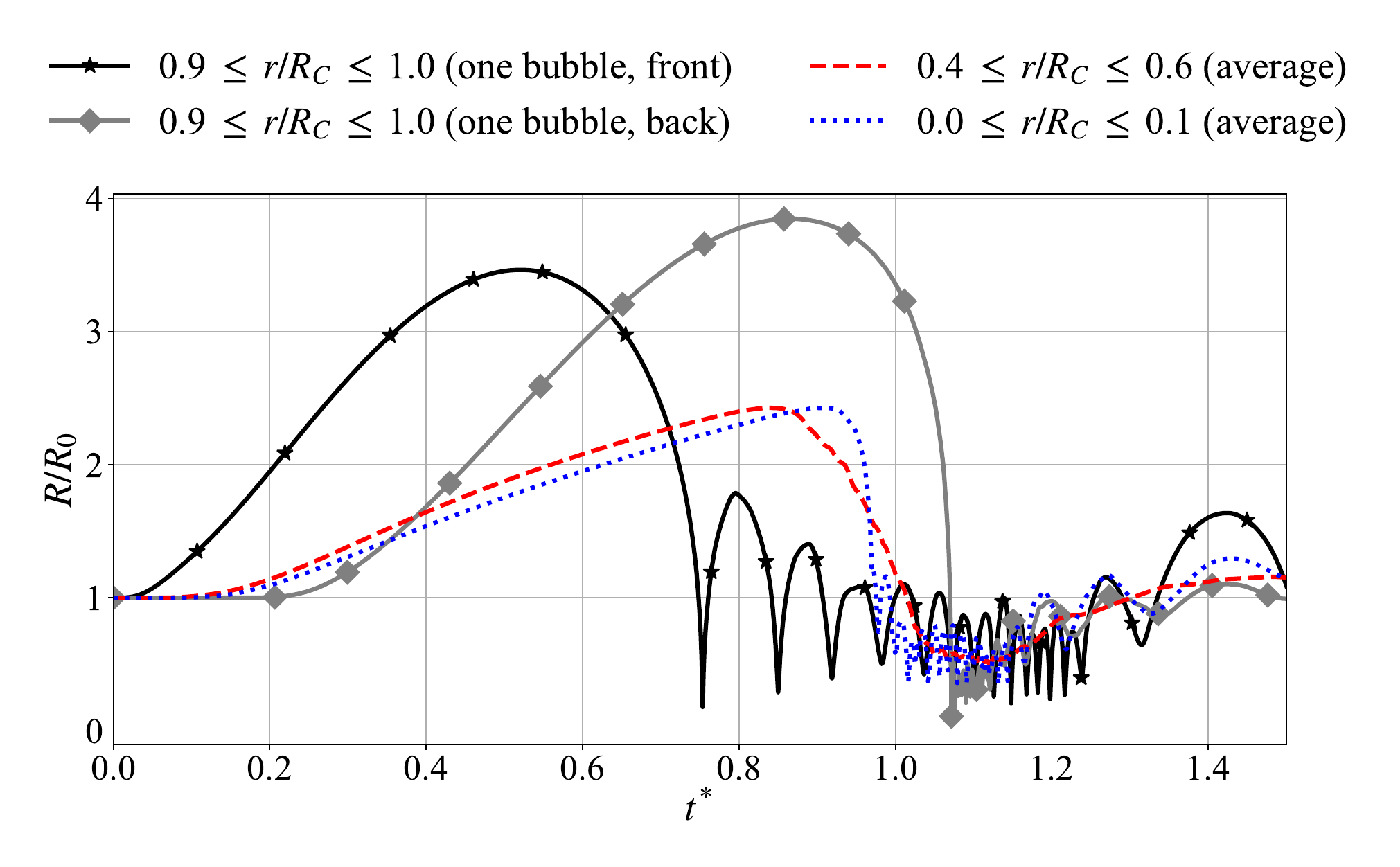}
    \caption{Evolution of the normalized bubble radius as a function of the dimensionless time $t^{*}=tf$ in selected locations of the spherical bubble cluster shown in Figure \ref{fig:Maeda2019_ClusterEvolution}, where $r$ refers to the radius with respect to the center of the cluster.}
    \label{fig:Maeda2019_RadiiEvolution}
\end{figure*}

\subsection{Asymmetric collapse of a spherical bubble cluster}
\label{sec:Maeda2019}

When a traveling acoustic waves interacts with a bubble cluster, the response of the individual bubbles depends on their location \citep{DAgostino1989,Wang1999,Arora2007}; bubbles in the center are shielded by their neighbors, thus peripheral bubbles respond stronger to the acoustic excitation than inner bubbles. 
Using three-dimensional compressible flow simulations in which the bubbles are described using the Keller-Miksis model in conjunction with a Lagrangian particle tracking algorithm, \citet{Maeda2019} reproduced this behavior for a monodisperse spherical bubble cluster with 2500 bubbles.

This test case involves a monodisperse spherical bubble cluster with a radius of $R_\mathrm{C} = 2.5 \ \mathrm{mm}$, containing $N = 2500$ bubbles of initial radius $R_{0} = 10 \ \upmu\mathrm{m}$. This cluster is excited by a single pulse of a sinusoidal wave propagating from left to right along the $x$-axis,
\begin{equation}
p_{\text{a}}(\mathbf{x},t) = \Delta p \, \sin \left( 2 \pi f \left[ t - \frac{x}{c} \right] \right) ,
\end{equation}
with pressure amplitude $\Delta p = -2.5 \ \mathrm{bar}$ and frequency $f = 50.0 \ \mathrm{kHz}$, resulting in a wavelength ($\lambda = 30 \ \mathrm{mm}$) that is longer than $R_\mathrm{C}$. 
An air-water system is considered here, described by the same properties as presented in Section \ref{sec:Haghi2019}.

Figure \ref{fig:Maeda2019_ClusterEvolution} shows the bubble cluster at four distinct time instances. Qualitatively, the bubble cluster exhibits the same asymmetric expansion and collapse as previously observed in experiments \citep{Arora2007,Wang1999} and three-dimensional compressible flow simulations \citep{Maeda2019}. 
The evolution of the normalized bubble radius $R/R_{0}$ as a function of the dimensionless time $t^{*}=tf$ in selected locations is shown in Figure \ref{fig:Maeda2019_RadiiEvolution}. In agreement with previously reported results \citep{Arora2007,Maeda2017}, the bubbles near the edge of the cluster respond stronger to the passing excitation wave than the bubbles located near the center of the bubble cluster.
This tendency is a consequence of the fact that bubbles near the edge of the bubble cluster have in general fewer immediate neighbors compared to the bubbles located in the interior of the bubble cluster, leading to fewer bubble-bubble interactions. 
In reality, the bubbles also disperse the incident wave and dissipate parts of its energy \citep{Wang1999, Ohl2024}, further exacerbating the observed tendencies.

\subsection{Onset of cavitation of interacting bubbles}
\label{sec:Ida2009}

The onset of cavitation signified by an explosive expansion of preexisting bubbles in response to a sudden reduction in ambient pressure has previously been shown to be influenced significantly by bubble-bubble interactions \citep{Ida2009, Fuster2014a}. Similar to the behavior observed with respect to the frequency response of multi-bubble systems in Sec.~\ref{sec:Haghi2019}, the behavior and response of larger bubbles dominate the behavior of smaller bubbles in their vicinity, which may delay the onset of cavitation of the smaller bubbles \citep{Ida2009}.

To study the onset of cavitation, the change in ambient pressure is defined, following the work of \citet{Ida2009}, as
\begin{equation}
    p_{\text{a}}(t) = \begin{cases}
        0 & t < T \\
        \dfrac{1 - \cos\left[ \dfrac{\pi}{T} (t+T) \right]}{2} \left( p_\mathrm{ng} - p_{0} \right) & T \leq t \leq 2T \\
        p_\mathrm{ng} - p_{0} & t > 2T, \label{eq:pa_Ida2009}
    \end{cases}
\end{equation}
with $T = 10.0 \ \upmu \mathrm{s}$ and $p_\mathrm{ng}$ being a desired negative pressure value. When neglecting the viscosity of the liquid, 
the liquid pressure at the bubble wall, Eq.~\eqref{eq:pL}, reduces to
\begin{equation}
    p_\mathrm{L} = p_\mathrm{G,0} \left( \frac{R_{0}^{3}}{R^{3}} \right)^{\kappa} - \frac{2 \sigma}{R} = \left( p_{0} + \frac{2 \sigma}{R_{0}} \right) \left( \frac{R_{0}^{3}}{R^{3}} \right)^{\kappa} - \frac{2 \sigma}{R}.
\end{equation}
Assuming the compression and expansion of the gas in the bubble is an isothermal process with $\kappa=1$, the critical pressure $p_{\mathrm{C}}$ for the onset of cavitation of a single bubble can be computed by solving $\mathrm{\partial}p_\mathrm{L} / \mathrm{\partial}R = 0$, leading to \citep{Neppiras1951}
\begin{equation}
    p_\mathrm{C} = -\sqrt{\dfrac{32 \sigma^{3}}{27 \left( p_{0} + \dfrac{2 \sigma}{R_{0}} \right) R_{0}^{3}}}.
    \label{eq:pc_Blake}
\end{equation}

Following the work of \citet{Ida2009}, we consider the response of a two-bubble system to a reduction in pressure from $p_0$ to $p_\mathrm{ng}$ as described by Eq.~\eqref{eq:pa_Ida2009}. The two bubbles are located at a distance of $\Delta x_{12}$ from each other and have an initial radius of $R_{1,0}= 2.0 \ \upmu \mathrm{m}$ and  $R_{2,0} = 20.0 \ \upmu \mathrm{m}$, respectively.
The system considered here is an air-water system, with a liquid density of $\rho = 1000 \ \mathrm{kg}/\mathrm{m}^{3}$, a speed of sound of $c = 1500 \ \mathrm{m}/\mathrm{s}$, a liquid viscosity of $\mu = 1.002 \times 10^{-3} \ \mathrm{Pa} \, \mathrm{s}$ and a polytropic exponent of the gas of $\kappa = 1$. 
The surface tension coefficient is $\sigma = 0.0728 \ \mathrm{N}/\mathrm{m}$ and the ambient pressure is $p_{0} = 1.013 \times 10^{5} \ \mathrm{Pa}$. It is worth noting that \citet{Ida2009} used a Rayleigh-Plesset equation with interaction terms treating the liquid as incompressible, given for bubble $i$ as
\begin{equation}
    R_i \ddot{R}_i + \frac{3}{2}\dot{R}^2_i = \frac{p_{\mathrm{L},i}-p_0-p_\mathrm{a}(t)}{\rho} - \sum_{j=1}^N \frac{1}{\Delta x_{12}} \frac{\mathrm{d}(R^2_j \dot{R}_j)}{\mathrm{d}t},
\end{equation}
which does not account for the finite propagation speed of the emitted acoustic waves.

\begin{figure*}
    \centering
    \includegraphics[width=0.75\textwidth]{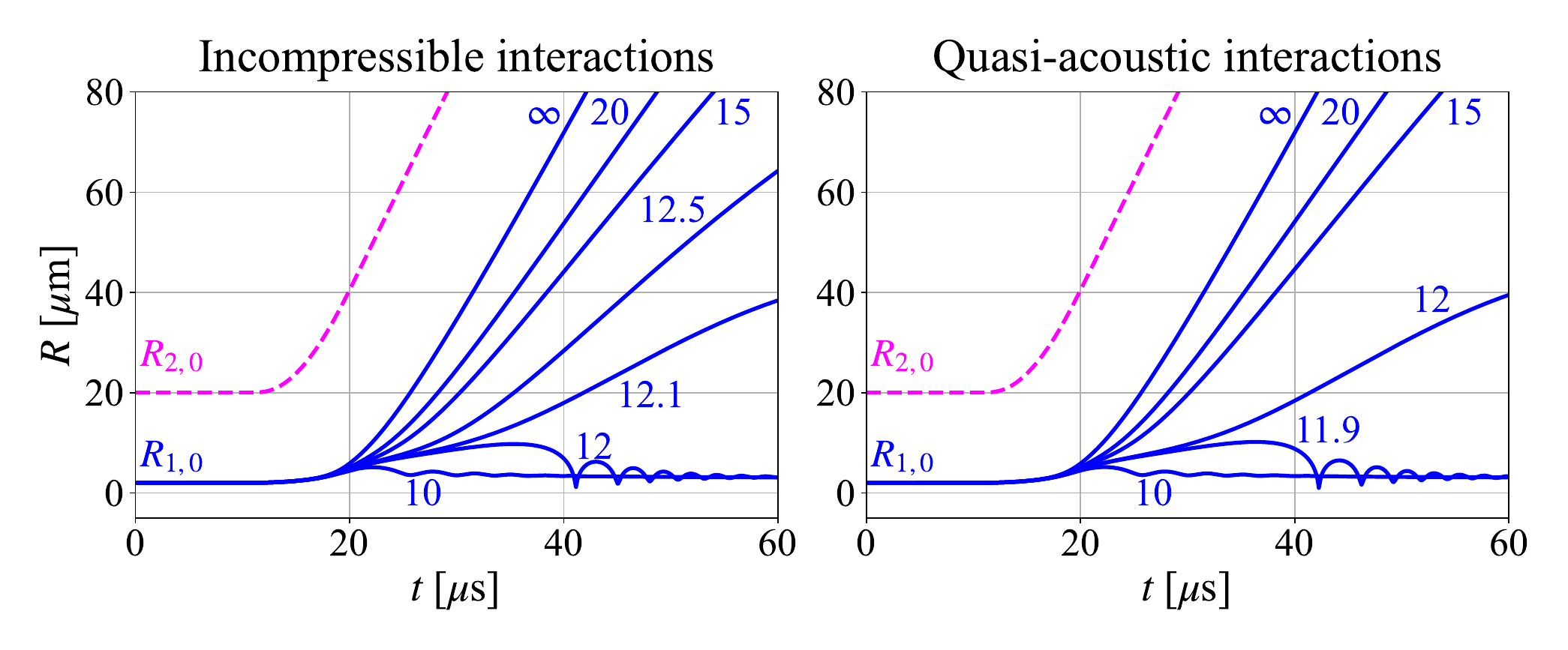}
    \caption{Evolution of the bubble radii of a two-bubble cluster in an air-water system with initial bubble radii $R_{1,0} = 2.0 \ \upmu\mathrm{m}$ (blue solid lines) and $R_{2,0} = 20.0 \ \upmu\mathrm{m}$ (pink dashed line), subject to a reduction in pressure described by Eq.~\eqref{eq:pa_Ida2009} from $p_0$ to $p_\mathrm{ng}=-0.25 \, p_0$, considering incompressible and quasi-acoustic interactions. The number next to each curve represents the dimensionless bubble-bubble distance, $\Delta x_{12} / \left( R_{1,0} + R_{2,0} \right)$.}
    \label{fig:Ida2009_CavitationInception_D_2Bubbles}
\end{figure*}

\begin{figure*}
    \centering
    \includegraphics[width=0.75\textwidth]{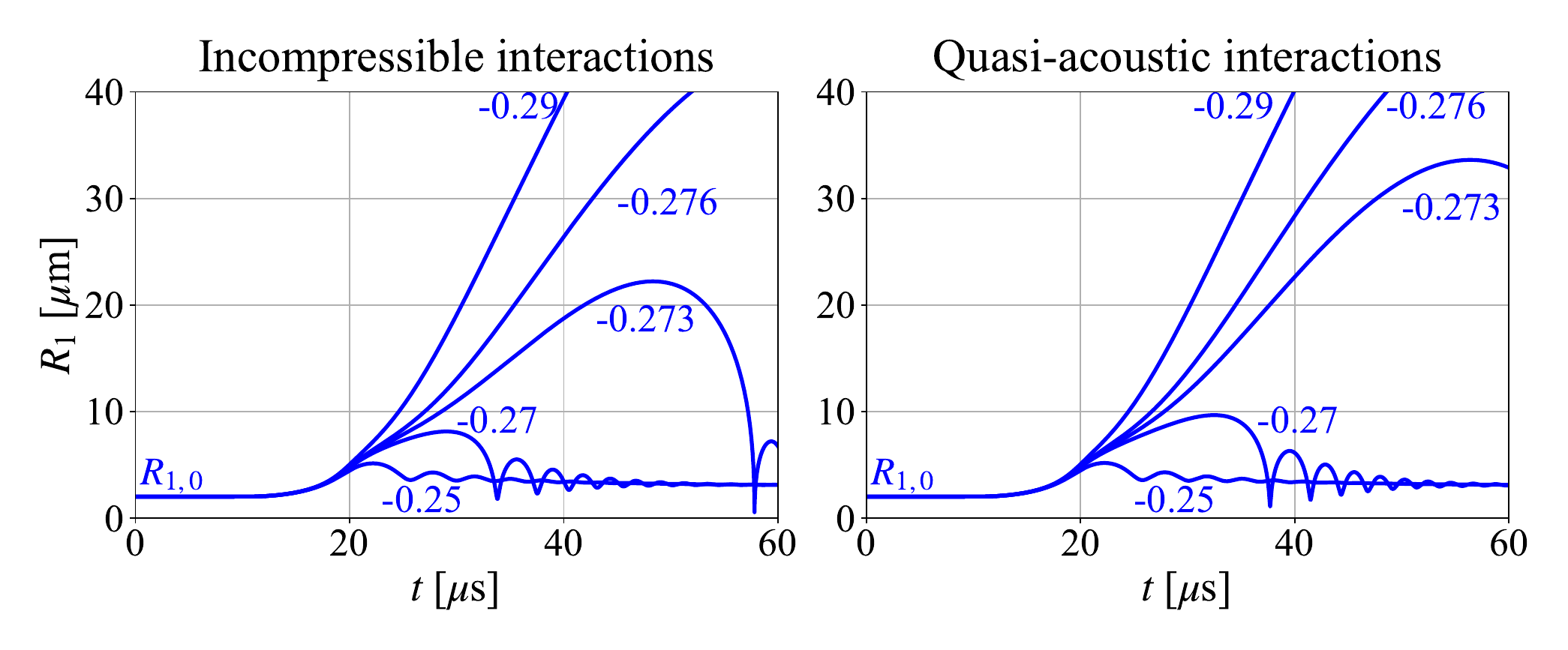}
    \caption{Evolution of the bubble radius of the smaller bubble with an initial radius of $R_{1,0} = 2.0 \ \upmu\mathrm{m}$ in the two-bubble cluster of Figure \ref{fig:Ida2009_CavitationInception_D_2Bubbles}, subject to a reduction in pressure described by Eq.~\eqref{eq:pa_Ida2009} from $p_0$ to $p_\mathrm{ng}$, considering incompressible and quasi-acoustic interactions. The dimensionless bubble-bubble distance is $\Delta x_{12} / \left( R_{1,0} + R_{2,0} \right) = 10$ and the number next to each  curve represents the dimensionless negative pressure $p_\mathrm{ng}/p_{0}$.}
    \label{fig:Ida2009_CavitationInception_Png_2Bubbles}
\end{figure*}

\begin{figure}
    \centering
    \includegraphics[width=\linewidth]{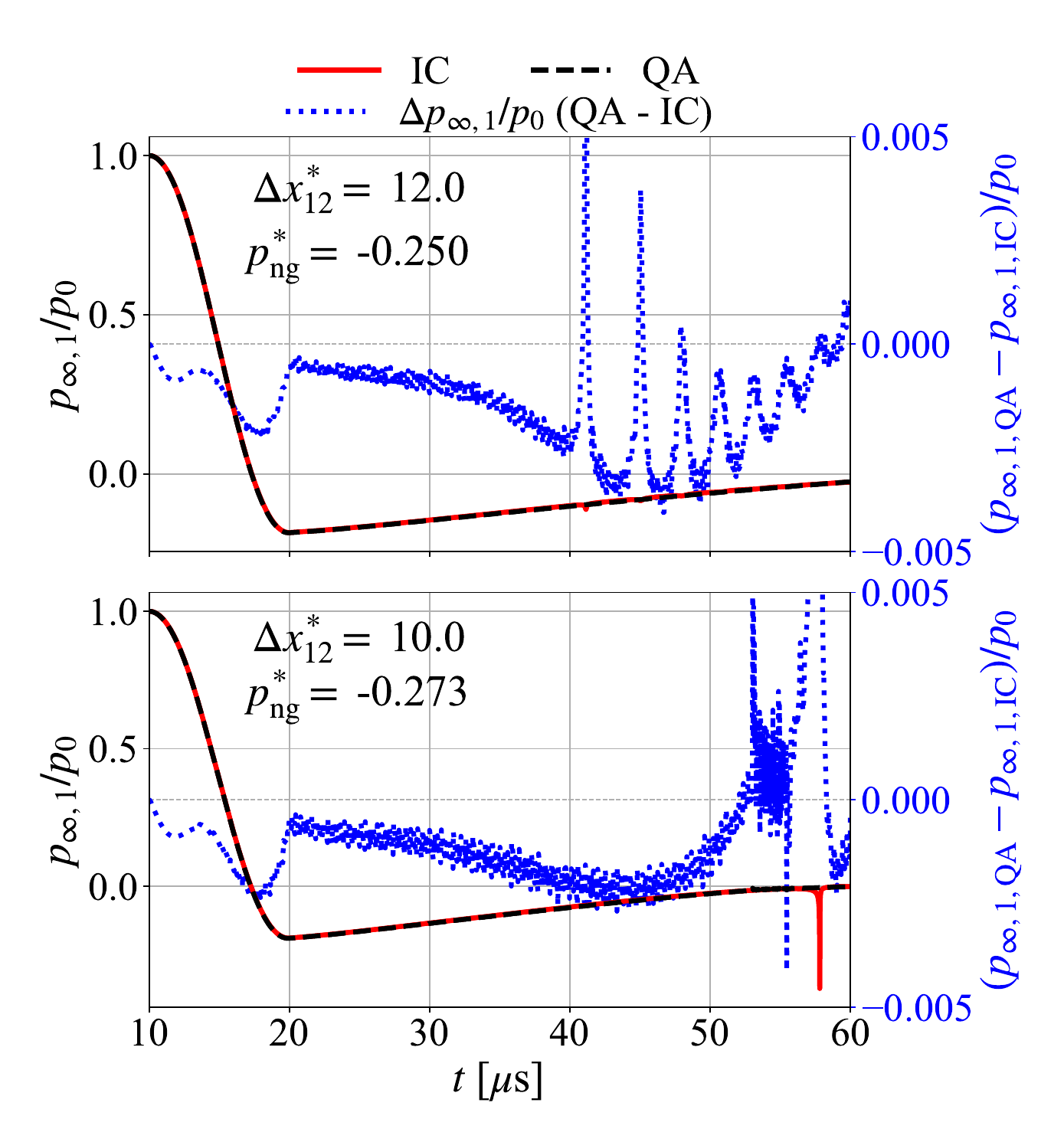}
    \caption{Evolution of the ambient pressure of the smaller bubble with an initial radius of $R_{1,0} = 2.0 \ \upmu\mathrm{m}$ in the two-bubble cluster of Figure \ref{fig:Ida2009_CavitationInception_D_2Bubbles} (left axis), subject to a reduction in pressure described by Eq.~\eqref{eq:pa_Ida2009} from $p_0$ to $p_\mathrm{ng}$, considering incompressible (IC) and quasi-acoustic (QA) interactions. The pressure difference between the two models is displayed on the right axis. Two parameter pairs of normalized dimensionless bubble-bubble distance $\Delta x_{12}^{*} = \Delta x_{12} / \left( R_{1,0} + R_{2,0} \right)$ and normalized negative pressure $p_{\mathrm{ng}}^{*} = p_\mathrm{ng}/p_{0}$ are considered.}
    \label{fig:Ida2009_CavitationInception_Png_2Bubbles_Pressure}
\end{figure}

\begin{figure}
    \centering
    \centerline{\includegraphics[width=1.1\linewidth]{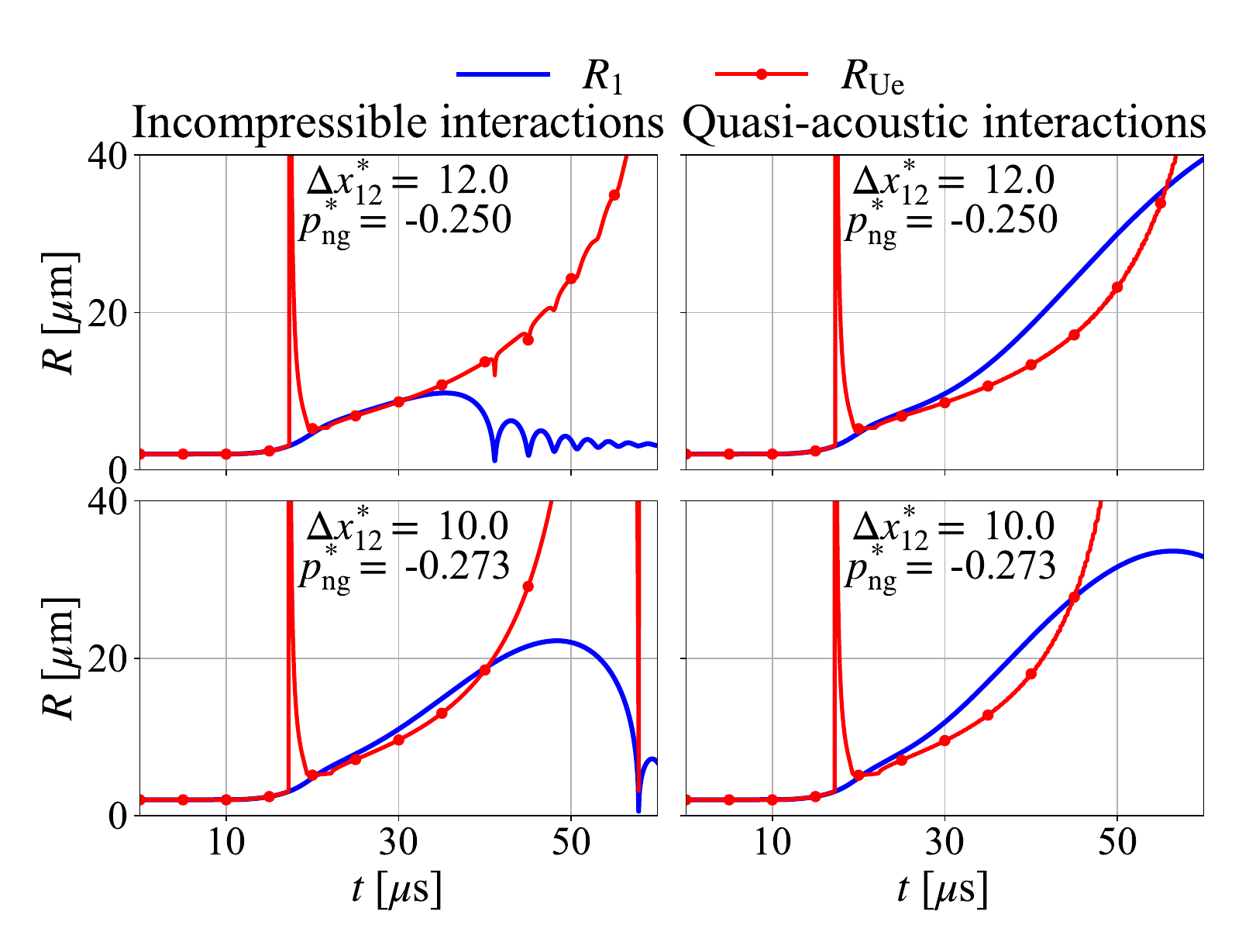}}
    \caption{Evolution of the bubble radius and the unstable equilibrium radius, obtained by solving Eq. \eqref{eq:R_unstable}, of the smaller bubble with an initial radius of $R_{1,0} = 2.0 \ \upmu\mathrm{m}$ in the two-bubble cluster of Figure \ref{fig:Ida2009_CavitationInception_D_2Bubbles}, subject to a reduction in pressure described by Eq.~\eqref{eq:pa_Ida2009} from $p_0$ to $p_\mathrm{ng}$, considering incompressible and quasi-acoustic interactions. The parameter pairs $\left( \Delta x_{12}^{*}, p_{\mathrm{ng}}^{*} \right)$ are the same as in Figure \ref{fig:Ida2009_CavitationInception_Png_2Bubbles_Pressure}.}
    \label{fig:Ida2009_CavitationInception_Png_2Bubbles_Unstableradius}
\end{figure}

\begin{figure*}
    \centering
    \includegraphics[width=0.75\textwidth]{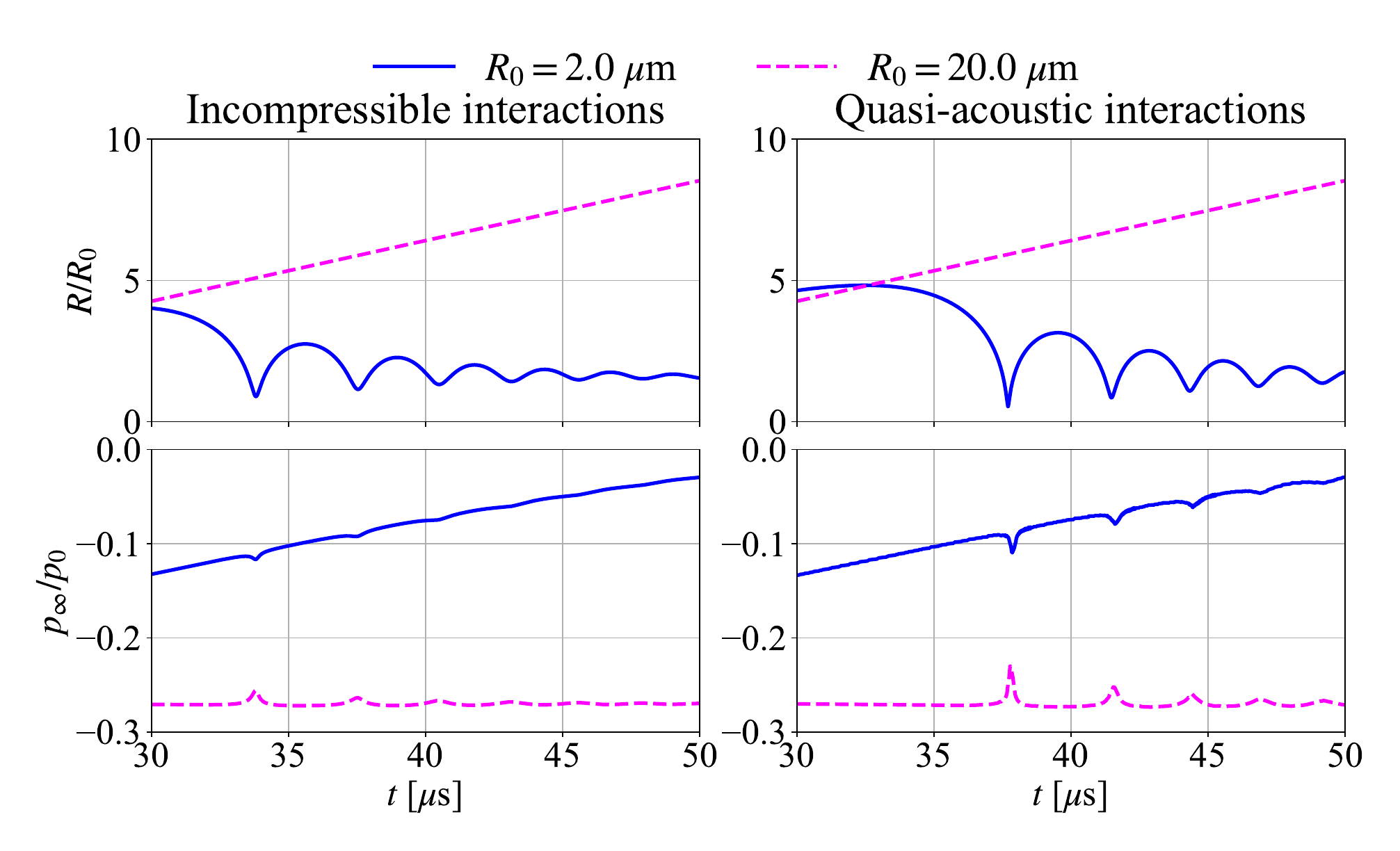}
    \caption{Evolution of the normalized bubble radii (top) and the normalized ambient pressure (bottom) in the two-bubble cluster of Figure \ref{fig:Ida2009_CavitationInception_D_2Bubbles}, subject to a reduction in pressure described by Eq.~\eqref{eq:pa_Ida2009} from $p_0$ to $p_\mathrm{ng}=-0.27 \, p_0$, considering incompressible and quasi-acoustic interactions. The dimensionless bubble-bubble distance is $\Delta x_{12} / \left( R_{1,0} + R_{2,0} \right) = 10$.}
    \label{fig:Ida2009_CavitationInception_Png_2Bubbles_Collapse}
\end{figure*}

Based on Eq.~\eqref{eq:pc_Blake}, the critical pressure for the onset of cavitation of the two considered bubbles is $p_{\mathrm{C},1} = - 0.179 \, p_{0}$ and $p_{\mathrm{C},2} = - 0.007 \, p_{0}$, respectively. As suggested by these critical pressure values, a negative pressure sufficient to promote the onset of cavitation of bubble 1 (the smaller bubble)
is also sufficient to promote the onset of cavitation of bubble 2 (the larger bubble).
Considering both incompressible and quasi-acoustic interactions,
Figure \ref{fig:Ida2009_CavitationInception_D_2Bubbles} shows the radius evolution of both bubbles in response to a reduction in pressure to $p_\mathrm{ng}/p_0=-0.25$, for different separation distances of the bubbles, and Figure \ref{fig:Ida2009_CavitationInception_Png_2Bubbles} shows the radius evolution of the smaller bubble (bubble 1) located at a distance of $\Delta x_{12}= 10 (R_{1,0} + R_{2,0})$, for different negative pressure values $p_\mathrm{ng}$.
The results depicted in Figures \ref{fig:Ida2009_CavitationInception_D_2Bubbles} and \ref{fig:Ida2009_CavitationInception_Png_2Bubbles} obtained with incompressible interactions are in very good 
agreement
with the results reported by \citet{Ida2009}. It is noticeable that the onset of cavitation for the smaller bubble is observed at a smaller bubble-bubble distance, 
see Figure
\ref{fig:Ida2009_CavitationInception_D_2Bubbles}, and that the smaller bubble expands more under the same pressure conditions, see Figure \ref{fig:Ida2009_CavitationInception_Png_2Bubbles}, when considering quasi-acoustic interactions compared to incompressible interactions.

To better understand the differences between the two interaction models, the ambient pressure evolution for the smaller bubble is considered, as seen with Figure \ref{fig:Ida2009_CavitationInception_Png_2Bubbles_Pressure} for two different parameter pairs $\left( \Delta x_{12}, p_{\mathrm{ng}} \right)$. After the pressure reduction defined by Eq. \eqref{eq:pa_Ida2009}, the ambient pressure felt by the smaller bubble increases due to the pressure waves radiated by the larger bubble. During the expansion phase of the smaller bubble, $p_{\infty,1,\mathrm{QA}} < p_{\infty,1,\mathrm{IC}}$, a consequence of the delayed interactions in the quasi-acoustic model.

For a bubble to continue expanding, it is necessary that $R(t) > R_{\mathrm{Ue}}(t)$,\citep{Ida2009} with $R_{\mathrm{Ue}}$ denoting the unstable equilibrium radius. $R_{\mathrm{Ue}}$ is obtained by solving
\begin{equation}
    p_{\infty}(t) = p_{\mathrm{G}0} \left( \frac{R_{0}}{R_{\mathrm{Ue}}(t)} \right)^{3 \kappa} - \frac{2 \sigma}{R_{\mathrm{Ue}}(t)}.
    \label{eq:R_unstable}
\end{equation}
Because $\kappa = 1$, $R_{\mathrm{Ue}}(t)$ is here the only real root of a third-order polynom. The smaller the ambient pressure, the smaller is the unstable radius. Because the ambient pressure is smaller when considering the quasi-acoustic model, it results in a smaller unstable radius, allowing the smaller bubble to expand more and for a longer time than in the case of incompressible interactions, see Figure \ref{fig:Ida2009_CavitationInception_Png_2Bubbles_Unstableradius}.

Lastly, we consider the evolution of the ambient pressure during the collapse phase, see Figure \ref{fig:Ida2009_CavitationInception_Png_2Bubbles_Collapse}. With both interaction models, during the collapse of the smaller bubble, the larger bubble sees an increase in its ambient pressure (dashed pink lines), while the smaller bubble experiences a pressure drop (solid blue lines). When collapsing, a bubble emits strong pressure waves that increase the ambient pressure felt by its neighbors for a short amount of time, hindering them in their expansion. This results in a smaller radiated pressure by the neighbor bubbles, explaining the ambient pressure drop experienced by the collapsing bubble. The stronger the collapse, the larger the pressure drop is, see the solid red curve in the bottom graph of Figure \ref{fig:Ida2009_CavitationInception_Png_2Bubbles_Pressure}.

\subsection{Onset of cavitation of a spherical bubble cluster}
\label{sec:tension}

To extend the test case presented in the previous section, 
we consider a spherical bubble cluster, similar to the cluster previously considered by \citet{Maeda2019} and in Section \ref{sec:Maeda2019}, which is subjected to a single tension pulse, defined as
\begin{equation}
    p_{\text{a}}(t) = - (p_{0} - p_{1}) \sin^{2}{\left( \frac{\pi t}{\tau} \right)},
    \label{eq:pa_tension}
\end{equation}
where $p_{1}<0$ and $\tau$ is the duration of the pulse.
The spherical bubble cluster has a radius of $R_\mathrm{C}=232 \ \upmu\mathrm{m}$ and consists of $N=250$ bubbles.
Two initial radius distributions are 
considered: monodisperse bubbles with the initial radius $R_{0}= 2.0 \ \upmu\mathrm{m}$ and polydisperse bubbles with a log-normal size distribution.
This log-normal distribution is defined by $\ln\left( R_{0} / R_{0,\mathrm{ref}} \right) \sim N \left( \overline{m} = 0, \varsigma = 0.7 \right)$, where $R_{0,\mathrm{ref}} = 2.0 \ \upmu\mathrm{m}$ is the reference bubble size, $\overline{m}$ is the mean and $\varsigma$ is the standard deviation of the distribution, with the values of $\overline{m}$ and $\varsigma$ adopted from previous work \citep{Ando2011, Maeda2019}. 
We again assume an air-water system, with the same fluid properties as used in Section \ref{sec:Haghi2019}.
This choice of parameters 
leads to a critical pressure for the onset of cavitation of $p_{\text{C}} = -2.67 \times 10^{4} \ \mathrm{Pa}$ with respect to a single bubble.
Both bubble clusters are subject to a tension wave described by Eq.~\eqref{eq:pa_tension}, with 
$p_{1} = -3.0 \times 10^{4} \ \mathrm{Pa}$ and $\tau = 1.75 \ \upmu\mathrm{s}$.
With this duration $\tau$, the pressure pulse has a wavelength of $\lambda_\mathrm{a} > 10 R_\mathrm{C}$, such that the assumption that the pulse is dependent on time
but not space, see Eq.~\eqref{eq:pa_tension}, is justified.

\begin{figure*}
    \centering
    \includegraphics[width=\linewidth]{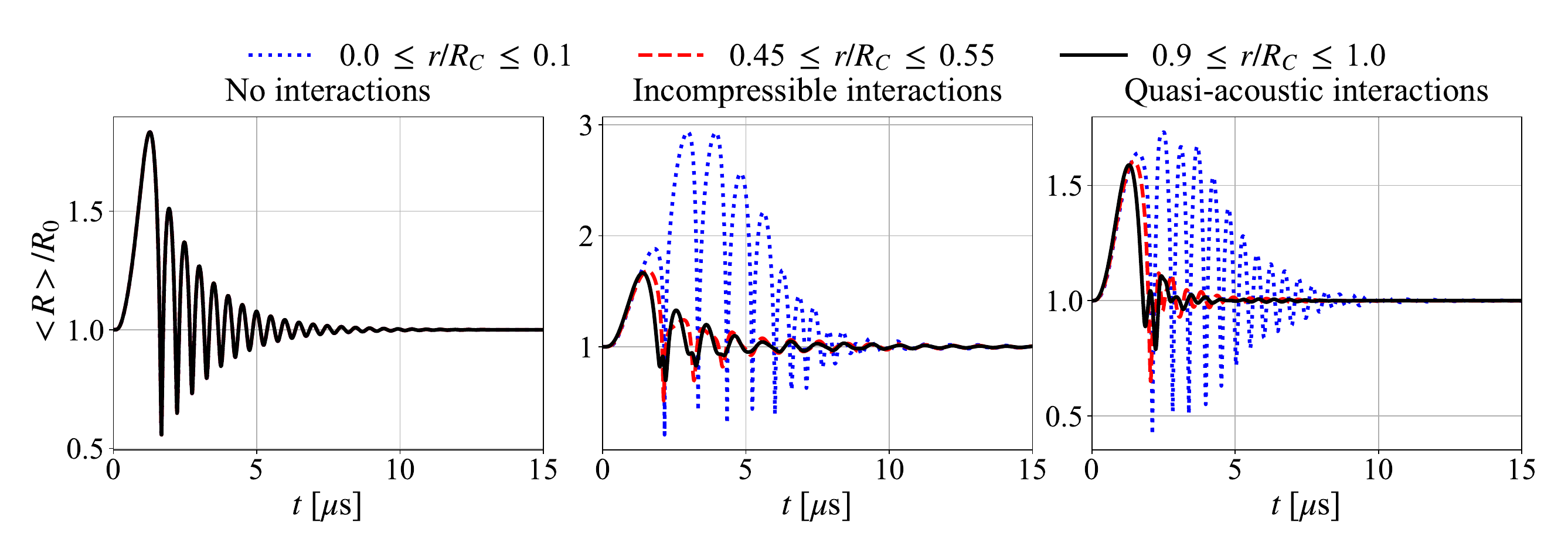}
    \caption{Evolution of the normalized mean bubble radius $<R>/R_{0}$ for the bubbles of a spherical monodisperse bubble cluster. \refereeone{Each line style represents the mean value for a specific group of bubbles depending on the radial location $r$ of the bubbles with respect to the center of the cluster (dashed for bubbles near the center, dotted for bubbles located at mid-distance and solid for bubbles at the edge of the cluster).} The cluster has a radius of $R_\mathrm{C}=232 \ \upmu\mathrm{m}$ and is composed of $250$ bubbles with an initial radius of $R_{0} = 2.0 \ \upmu\mathrm{m}$. The cluster is excited by a single tension pulse described by Eq. (39), with a duration of $\tau = 1.75 \ \upmu\mathrm{s}$ and a pressure amplitude of $p_{1} = -3.0 \times 10^{4} \ \mathrm{Pa}$.}
    \label{fig:tension_monodispersed}
\end{figure*}

\begin{figure*}
    \centering
    \includegraphics[width=\linewidth]{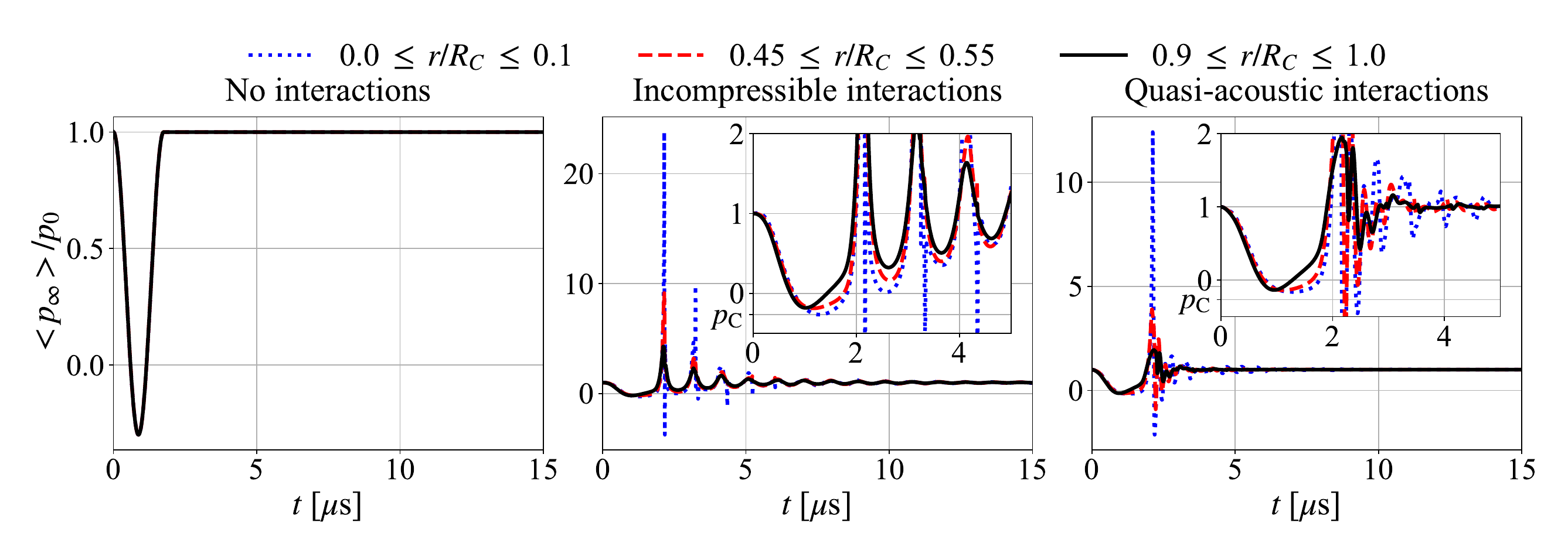}
    \caption{Evolution of the normalized mean pressure $<p_{\infty}>/p_{0}$ experienced by the bubbles of the spherical monodisperse bubble cluster shown in Figure \ref{fig:tension_monodispersed}, depending on the radial bubble location $r$ with respect to the center of the cluster.}
    \label{fig:tension_monodispersed_press}
\end{figure*}

\begin{figure*}
    \centering
    \includegraphics[width=\linewidth]{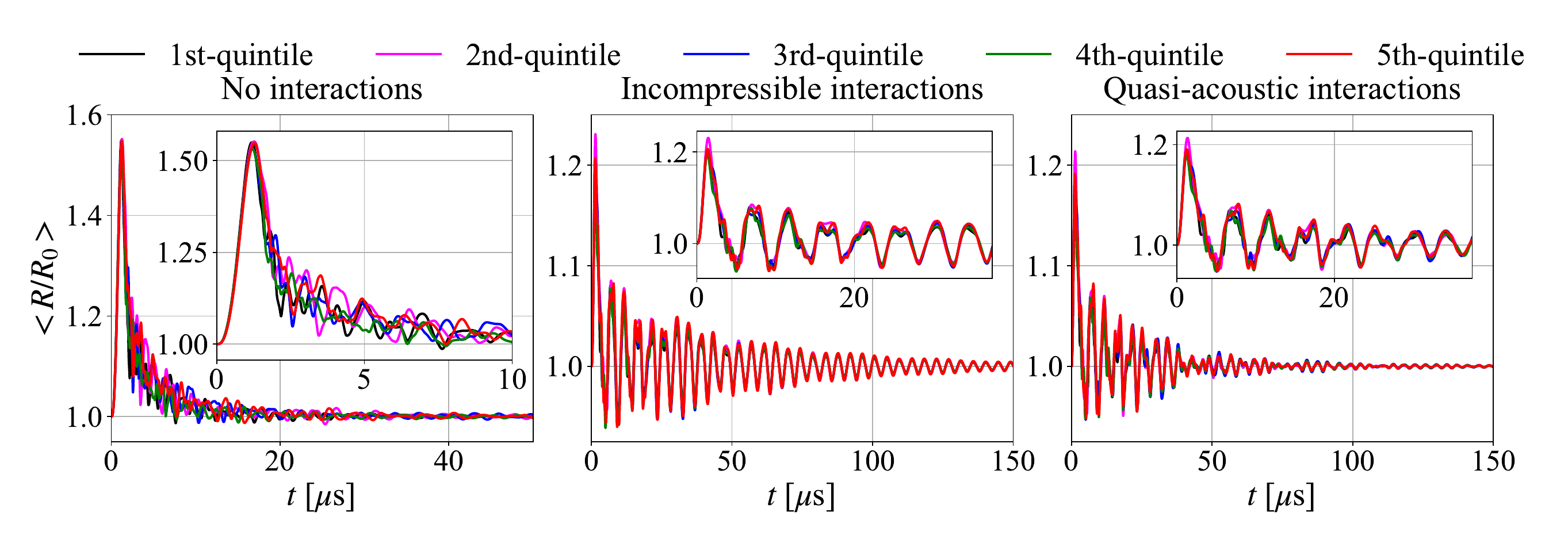}
    \caption{
    Evolution of the normalized mean bubble radius $<R/R_{0}>$ for the bubbles of a spherical polydisperse bubble cluster, depending on their initial radius $R_{0}$ in a quintile repartition of the initial radii distribution.
    The cluster has a radius of $R_\mathrm{C}=232 \ \upmu\mathrm{m}$ and is composed of $250$ bubbles, the size of which is described by a log-normal distribution with a initial reference radius of $R_{0,\mathrm{ref}} = 2.0 \ \upmu\mathrm{m}$ and a standard deviation of $\varsigma = 0.7$. The cluster is excited by a single tension pulse described by Eq.~\eqref{eq:pa_tension}, with a duration of $\tau = 1.75 \ \mu\mathrm{s}$ and a pressure amplitude of $p_{1} = -3.0 \times 10^{4} \ \mathrm{Pa}$.}
    \label{fig:tension_polydispersed}
\end{figure*}

The results for the monodisperse cluster are shown in Figures \ref{fig:tension_monodispersed} and \ref{fig:tension_monodispersed_press}, obtained with incompressible and quasi-acoustic interactions, as well as by neglecting interactions between the bubbles. 
When the interactions between the bubble are neglected, every bubble in the cluster experiences the same pressure, leading to the same radius evolution. 
Interestingly, with both interaction models, the response to the pulse of the bubbles near the cluster center is most pronounced.
Since the considered tension pulse is only dependent on time but not space,
the asymmetric collapse tendency highlighted by \citet{Wang1999} or \citet{Nasibullaeva2013}, and also observed in Section \ref{sec:Maeda2019}, is not present here. In the present case, the number of close neighbors is the key factor influencing the response of a bubble. Near the edge of the bubble cluster, the tension pulse 
is the dominant
excitation experienced by the bubbles since they have few close neighbors. Closer to the center of the bubble cluster, the number of neighbors for each bubble increases, and the pressure emitted by these neighbors becomes an additional excitation for these bubbles.
As seen in Figure \ref{fig:tension_monodispersed_press}, the magnitude of the normalized mean pressure experienced by the bubbles in a specific region of the bubble cluster is highest near the center of the cluster. 
The incompressible interactions yield larger amplitudes of the radius and pressure oscillations than the quasi-acoustic interactions.
In fact, when neglecting the propagation time of the emission, the pressure changes induced by the interactions are felt at the same time for each bubble in the cluster, resulting in pressure peaks that are larger and more concentrated in time. The incompressible interactions also maintain the excitation state initiated by the incident pulse for a longer time, due to the larger and faster initial pressure increase. Indeed, this first pressure peak is followed by two smaller peaks, as seen in the center graph of Figure \ref{fig:tension_monodispersed_press}. Such secondary pressure peaks are not observed when considering quasi-acoustic interactions, see the right graph of Figure \ref{fig:tension_monodispersed_press}, explaining the faster recovery of the undisturbed state.

The results of the polydisperse cluster are presented in Figure \ref{fig:tension_polydispersed}, showing the dimensionless mean radius based on a quintile repartition of the initial radius distribution. 
Neglecting the interactions between the bubbles, small discrepancies between the dynamics of each bubble group are noticeable. However these differences in behavior disappear when taking bubble-bubble interactions into account. 
This tendency is consistent with previous numerical observations reported by \citet{Nasibullaeva2013}, where in a polydisperse cluster with two initial bubble radii groups, both bubble groups synchronize their oscillations, similar to what is seen in the Figure \ref{fig:tension_polydispersed}. 
Also, each bubble attains a smaller maximum radius when bubble-bubble interactions are taken into account, as a results of the acoustic emissions of their neighbors.
The main difference in this case is between the two interaction models: when considering quasi-acoustic interactions, the system is more damped compared to when incompressible interactions are considered, with a faster decay of the oscillation amplitude for each bubble group. Like stated with the monodisperse cluster, the delayed interactions maintain for a shorter time the excited state induced by the incident pulse.

\section{Conclusions}
\label{sec:conclusions}

We have presented a new model for the bubble dynamics, acoustic emissions and interactions based on the quasi-acoustic assumption and in spherical symmetry. This model builds on the Keller-Miksis equation for the radial bubble dynamics and a Lagrangian wave tracking approach of the acoustic emissions and interactions of the bubbles, and takes the compressibility of the liquid surrounding the bubbles consistently into account up to first order in the Mach number.  

Representative and well-defined test-cases, including monodisperse and polydisperse bubble clusters, have been used to validate the proposed model and highlight the differences compared to the commonly used class of models derived under the assumption of an incompressible liquid.
The presented results show excellent agreement with previously reported phenomena of multi-bubble systems, such as the dominance of larger bubbles on the resonance behavior of smaller bubbles, the asymmetric collapse of bubble clusters, and the delayed onset of cavitation of smaller bubbles in the vicinity of larger bubbles. 
Contrary to the widely used models based on incompressible interactions, accounting for the finite propagation speed of the acoustic emissions, has been found to affect the oscillation patterns in bubble screens, the onset of cavitation of multi-bubble systems, as well as the response of bubble clusters to tension waves. \refereeone{The differences between models based on incompressible interactions and the proposed quasi-acoustic model are 
most
pronounced for large and dense bubble systems,
in which the finite propagation speed  and the resulting time delay of the acoustic interactions becomes appreciable 
and results in markedly different bubble dynamics.}

In deriving and testing the proposed model based on the quasi-acoustic assumption, we have neglected the translational motion of the bubbles, for instance as a result of secondary Bjerknes forces associated with the acoustic interaction of the bubbles. However, because the proposed quasi-acoustic model readily provides the flow velocity and pressure associated with the acoustic emissions of the bubbles, see Eqs.~\eqref{eq:u_inter} and \eqref{eq:dp_x}, combining the proposed model with a methodology to track the motion of the bubbles would be straightforward. \refereeone{Additionally, the use of the Keller-Miksis equation limits this model to cases 
with moderate pressure amplitudes
such that the 
density and the speed of sound remain approximately constant. For water, this assumption yields errors with respect to the density and speed of sound that are smaller than $6\ \%$ for pressures up to $50 \, \text{MPa}$ compared to ambient conditions, according to the IAPWS R6-95(2018) standard \citep{Wagner2002}.}

In summary, the proposed model provides a consistent computational tool to simulate the radial dynamics, acoustic emissions and interactions of multi-bubble systems, enabling a more faithful prediction of the acoustic emissions and interactions than the frequently employed incompressible models, \refereefour{with a reduced complexity and faster execution compared to state-of-the-art CFD methods.} 

\section*{Acknowledgements}
We acknowledge the support of the Natural Sciences and Engineering Research Council of Canada (NSERC), funding reference number RGPIN-2024-04805, as well as fruitful discussions with Yuzhe Fan, Daniel Fuster, Hossein Haghi and Sören Schenke on acoustic bubble-bubble interactions and the numerical modeling of cavitation.

\section*{Data availability}
Data supporting this study is available at \url{https://doi.org/10.5281/zenodo.13891010} under a Creative Commons Attribution license.

\appendix

\section{Computing the interaction terms}
\label{sec:computinginteractionterms}

\begin{figure*}
    \centering
    \includegraphics[scale=1.0]{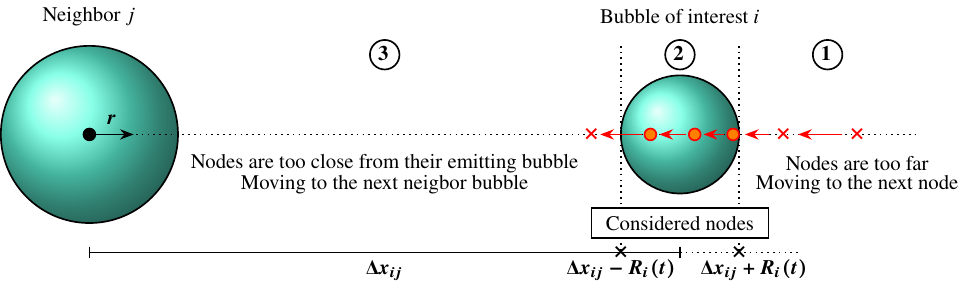}
    \caption{Schematic illustration of computing the interactions between two bubbles using the quasi-acoustic model. To compute the invariants $\phi_j$ and $g_j$ required for computing the interaction pressure $p_\mathrm{inter}$ in Eq.~\eqref{eq:pinf_bubble} for the bubble of interest $i$, the list of emission nodes of neighbor bubble $j$ is parsed to identify all emission nodes that coincide with bubble $i$.}
    \label{fig:bubbles_interacting}
\end{figure*}

The quasi-acoustic interactions are modeled using the Lagrangian wave tracking previously proposed by \citet{Denner2023}. The information carried by the Lagrangian wave tracking allows to compute the interaction terms given in Eq.~\eqref{eq:pinf_bubble} for the neighbor bubbles. For each bubble, $\phi$ and $g$ are obtained from their neighbor bubbles at predefined time intervals by employing the following strategy, illustrated in Figure \ref{fig:bubbles_interacting}. Considering the bubble of interest $i$ and its neighbor $j$, we commence by parsing the list of emission nodes of bubble $j$, starting with the emission node that is furthest away from its bubble $j$. When all emission nodes of bubble $j$ that are located inside bubble $i$ are identified, we stop parsing the list of emission nodes of bubble $j$. The mean values of the invariants $\phi_j$ and $g_j$ of all emission nodes of bubble $j$ located inside bubble $i$ are then used to compute the interaction pressure $p_\mathrm{inter}$ in Eq.~\eqref{eq:pinf_bubble}. 
Other ways of determining $\phi_j$ and $g_j$ may also be considered, such as weighting the contribution of the emission nodes by a Gaussian or a radial basis function with finite support. The time derivative $\dot{p}_\mathrm{int}$ of the resulting interaction pressure, which is part of the pressure derivative $\dot{p}_\infty$ in Eq.~\eqref{eq:km_bubble}, is computed numerically as
\begin{equation}
        \dot{p}_{\text{inter}}\left( t_{i} \right) \approx \frac{p_{\text{inter}} \left(t_{i}\right) - p_{\text{inter}} \left(t_{i-1}\right)}{t_{i} - t_{i-1}}.
\end{equation}

To reduce the computational cost for large bubble clusters,
we prune emission nodes that do not carry relevant information.
Based on Eq.~\eqref{eq:p_general}, an emission node is considered to be obsolete if its pressure amplitude is
\begin{equation}
    |p(r,t) - p_\infty(t)| = \left| \rho_0 \left[ \frac{g(\tau)}{r(t)} -  \dfrac{u(r,t)^2}{2} \right] \right| < C \, \frac{p_\infty(t)}{N} .
    \label{eq:pruning}
\end{equation}
The rationale behind this condition is that emission nodes make a significant contribution to the interactions only if the global pressure considering every node is relevant in comparison to $p_{\infty}$. Since at the moment we consider every bubble to be interacting with all bubbles of a cluster, $p_{\infty}$ is divided by the total number of bubbles $N$. 
The coefficient $C$ is a case-dependent coefficient that has so far been determined mainly by trial and error, 
with the goal of reducing the execution time without neglecting the main pressure changes induced by the interactions and thereby compromising the results.


\begin{thebibliography}{57}%
    \makeatletter
    \providecommand \@ifxundefined [1]{%
     \@ifx{#1\undefined}
    }%
    \providecommand \@ifnum [1]{%
     \ifnum #1\expandafter \@firstoftwo
     \else \expandafter \@secondoftwo
     \fi
    }%
    \providecommand \@ifx [1]{%
     \ifx #1\expandafter \@firstoftwo
     \else \expandafter \@secondoftwo
     \fi
    }%
    \providecommand \natexlab [1]{#1}%
    \providecommand \enquote  [1]{``#1''}%
    \providecommand \bibnamefont  [1]{#1}%
    \providecommand \bibfnamefont [1]{#1}%
    \providecommand \citenamefont [1]{#1}%
    \providecommand \href@noop [0]{\@secondoftwo}%
    \providecommand \href [0]{\begingroup \@sanitize@url \@href}%
    \providecommand \@href[1]{\@@startlink{#1}\@@href}%
    \providecommand \@@href[1]{\endgroup#1\@@endlink}%
    \providecommand \@sanitize@url [0]{\catcode `\\12\catcode `\$12\catcode `\&12\catcode `\#12\catcode `\^12\catcode `\_12\catcode `\%12\relax}%
    \providecommand \@@startlink[1]{}%
    \providecommand \@@endlink[0]{}%
    \providecommand \url  [0]{\begingroup\@sanitize@url \@url }%
    \providecommand \@url [1]{\endgroup\@href {#1}{\urlprefix }}%
    \providecommand \urlprefix  [0]{URL }%
    \providecommand \Eprint [0]{\href }%
    \providecommand \doibase [0]{http://dx.doi.org/}%
    \providecommand \selectlanguage [0]{\@gobble}%
    \providecommand \bibinfo  [0]{\@secondoftwo}%
    \providecommand \bibfield  [0]{\@secondoftwo}%
    \providecommand \translation [1]{[#1]}%
    \providecommand \BibitemOpen [0]{}%
    \providecommand \bibitemStop [0]{}%
    \providecommand \bibitemNoStop [0]{.\EOS\space}%
    \providecommand \EOS [0]{\spacefactor3000\relax}%
    \providecommand \BibitemShut  [1]{\csname bibitem#1\endcsname}%
    \let\auto@bib@innerbib\@empty
    \bibitem [{\citenamefont {Matsumoto}\ and\ \citenamefont {Yoshizawa}(2005)}]{Matsumoto2005}%
      \BibitemOpen
      \bibfield  {author} {\bibinfo {author} {\bibfnamefont {Y.}~\bibnamefont {Matsumoto}}\ and\ \bibinfo {author} {\bibfnamefont {S.}~\bibnamefont {Yoshizawa}},\ }\bibfield  {title} {{\selectlanguage {en}\enquote {\bibinfo {title} {Behaviour of a bubble cluster in an ultrasound field},}\ }}\href {\doibase 10.1002/fld.833} {\bibfield  {journal} {\bibinfo  {journal} {International Journal for Numerical Methods in Fluids}\ }\textbf {\bibinfo {volume} {47}},\ \bibinfo {pages} {591--601} (\bibinfo {year} {2005})}\BibitemShut {NoStop}%
    \bibitem [{\citenamefont {Song}, \citenamefont {Moldovan},\ and\ \citenamefont {Prentice}(2019)}]{Song2019}%
      \BibitemOpen
      \bibfield  {author} {\bibinfo {author} {\bibfnamefont {J.~H.}\ \bibnamefont {Song}}, \bibinfo {author} {\bibfnamefont {A.}~\bibnamefont {Moldovan}}, \ and\ \bibinfo {author} {\bibfnamefont {P.}~\bibnamefont {Prentice}},\ }\bibfield  {title} {{\selectlanguage {en}\enquote {\bibinfo {title} {Non-linear {Acoustic} {Emissions} from {Therapeutically} {Driven} {Contrast} {Agent} {Microbubbles}},}\ }}\href {\doibase 10.1016/j.ultrasmedbio.2019.04.005} {\bibfield  {journal} {\bibinfo  {journal} {Ultrasound in Medicine \& Biology}\ }\textbf {\bibinfo {volume} {45}},\ \bibinfo {pages} {2188--2204} (\bibinfo {year} {2019})}\BibitemShut {NoStop}%
    \bibitem [{\citenamefont {Franc}\ and\ \citenamefont {Michel}(2005)}]{Franc2005}%
      \BibitemOpen
      \bibfield  {author} {\bibinfo {author} {\bibfnamefont {J.-P.}\ \bibnamefont {Franc}}\ and\ \bibinfo {author} {\bibfnamefont {J.-M.}\ \bibnamefont {Michel}},\ }\href {\doibase 10.1007/1-4020-2233-6} {{\selectlanguage {en}\emph {\bibinfo {title} {Fundamentals of {Cavitation}}}}},\ \bibinfo {series} {Fluid {Mechanics} and {Its} {Applications}}, Vol.~\bibinfo {volume} {76}\ (\bibinfo  {publisher} {Kluwer Academic Publishers},\ \bibinfo {address} {Dordrecht},\ \bibinfo {year} {2005})\BibitemShut {NoStop}%
    \bibitem [{\citenamefont {Hsiao}\ and\ \citenamefont {Chahine}(2015)}]{Hsiao2015}%
      \BibitemOpen
      \bibfield  {author} {\bibinfo {author} {\bibfnamefont {C.-T.}\ \bibnamefont {Hsiao}}\ and\ \bibinfo {author} {\bibfnamefont {G.~L.}\ \bibnamefont {Chahine}},\ }\bibfield  {title} {{\selectlanguage {en}\enquote {\bibinfo {title} {Dynamic response of a composite propeller blade subjected to shock and bubble pressure loading},}\ }}\href {\doibase 10.1016/j.jfluidstructs.2015.01.012} {\bibfield  {journal} {\bibinfo  {journal} {Journal of Fluids and Structures}\ }\textbf {\bibinfo {volume} {54}},\ \bibinfo {pages} {760--783} (\bibinfo {year} {2015})}\BibitemShut {NoStop}%
    \bibitem [{\citenamefont {Dular}\ \emph {et~al.}(2019)\citenamefont {Dular}, \citenamefont {Požar}, \citenamefont {Zevnik},\ and\ \citenamefont {Petkovšek}}]{Dular2019}%
      \BibitemOpen
      \bibfield  {author} {\bibinfo {author} {\bibfnamefont {M.}~\bibnamefont {Dular}}, \bibinfo {author} {\bibfnamefont {T.}~\bibnamefont {Požar}}, \bibinfo {author} {\bibfnamefont {J.}~\bibnamefont {Zevnik}}, \ and\ \bibinfo {author} {\bibfnamefont {R.}~\bibnamefont {Petkovšek}},\ }\bibfield  {title} {{\selectlanguage {en}\enquote {\bibinfo {title} {High speed observation of damage created by a collapse of a single cavitation bubble},}\ }}\href {\doibase 10.1016/j.wear.2018.11.004} {\bibfield  {journal} {\bibinfo  {journal} {Wear}\ }\textbf {\bibinfo {volume} {418-419}},\ \bibinfo {pages} {13--23} (\bibinfo {year} {2019})}\BibitemShut {NoStop}%
    \bibitem [{\citenamefont {Salzar}\ \emph {et~al.}(2017)\citenamefont {Salzar}, \citenamefont {Treichler}, \citenamefont {Wardlaw}, \citenamefont {Weiss},\ and\ \citenamefont {Goeller}}]{Salzar2017}%
      \BibitemOpen
      \bibfield  {author} {\bibinfo {author} {\bibfnamefont {R.~S.}\ \bibnamefont {Salzar}}, \bibinfo {author} {\bibfnamefont {D.}~\bibnamefont {Treichler}}, \bibinfo {author} {\bibfnamefont {A.}~\bibnamefont {Wardlaw}}, \bibinfo {author} {\bibfnamefont {G.}~\bibnamefont {Weiss}}, \ and\ \bibinfo {author} {\bibfnamefont {J.}~\bibnamefont {Goeller}},\ }\bibfield  {title} {{\selectlanguage {en}\enquote {\bibinfo {title} {Experimental {Investigation} of {Cavitation} as a {Possible} {Damage} {Mechanism} in {Blast}-{Induced} {Traumatic} {Brain} {Injury} in {Post}-{Mortem} {Human} {Subject} {Heads}},}\ }}\href {\doibase 10.1089/neu.2016.4600} {\bibfield  {journal} {\bibinfo  {journal} {Journal of Neurotrauma}\ }\textbf {\bibinfo {volume} {34}},\ \bibinfo {pages} {1589--1602} (\bibinfo {year} {2017})}\BibitemShut {NoStop}%
    \bibitem [{\citenamefont {Walls}\ \emph {et~al.}(2017)\citenamefont {Walls}, \citenamefont {McRae}, \citenamefont {Natarajan}, \citenamefont {Johnson}, \citenamefont {Antoniou},\ and\ \citenamefont {Bird}}]{Walls2017}%
      \BibitemOpen
      \bibfield  {author} {\bibinfo {author} {\bibfnamefont {P.~L.~L.}\ \bibnamefont {Walls}}, \bibinfo {author} {\bibfnamefont {O.}~\bibnamefont {McRae}}, \bibinfo {author} {\bibfnamefont {V.}~\bibnamefont {Natarajan}}, \bibinfo {author} {\bibfnamefont {C.}~\bibnamefont {Johnson}}, \bibinfo {author} {\bibfnamefont {C.}~\bibnamefont {Antoniou}}, \ and\ \bibinfo {author} {\bibfnamefont {J.~C.}\ \bibnamefont {Bird}},\ }\bibfield  {title} {{\selectlanguage {en}\enquote {\bibinfo {title} {Quantifying the potential for bursting bubbles to damage suspended cells},}\ }}\href {\doibase 10.1038/s41598-017-14531-5} {\bibfield  {journal} {\bibinfo  {journal} {Scientific Reports}\ }\textbf {\bibinfo {volume} {7}},\ \bibinfo {pages} {15102} (\bibinfo {year} {2017})}\BibitemShut {NoStop}%
    \bibitem [{\citenamefont {Johnsen}\ and\ \citenamefont {Colonius}(2009)}]{Johnsen2009}%
      \BibitemOpen
      \bibfield  {author} {\bibinfo {author} {\bibfnamefont {E.}~\bibnamefont {Johnsen}}\ and\ \bibinfo {author} {\bibfnamefont {T.}~\bibnamefont {Colonius}},\ }\bibfield  {title} {\enquote {\bibinfo {title} {Numerical simulations of non-spherical bubble collapse},}\ }\href {\doibase 10.1017/S0022112009006351} {\bibfield  {journal} {\bibinfo  {journal} {Journal of Fluid Mechanics}\ }\textbf {\bibinfo {volume} {629}},\ \bibinfo {pages} {231--262} (\bibinfo {year} {2009})}\BibitemShut {NoStop}%
    \bibitem [{\citenamefont {Ida}(2009)}]{Ida2009}%
      \BibitemOpen
      \bibfield  {author} {\bibinfo {author} {\bibfnamefont {M.}~\bibnamefont {Ida}},\ }\bibfield  {title} {{\selectlanguage {en}\enquote {\bibinfo {title} {Multibubble cavitation inception},}\ }}\href {\doibase 10.1063/1.3265547} {\bibfield  {journal} {\bibinfo  {journal} {Physics of Fluids}\ }\textbf {\bibinfo {volume} {21}},\ \bibinfo {pages} {113302} (\bibinfo {year} {2009})}\BibitemShut {NoStop}%
    \bibitem [{\citenamefont {Wang}\ and\ \citenamefont {Brennen}(1999)}]{Wang1999}%
      \BibitemOpen
      \bibfield  {author} {\bibinfo {author} {\bibfnamefont {Y.-C.}\ \bibnamefont {Wang}}\ and\ \bibinfo {author} {\bibfnamefont {C.~E.}\ \bibnamefont {Brennen}},\ }\bibfield  {title} {\enquote {\bibinfo {title} {Numerical {Computation} of {Shock} {Waves} in a {Spherical} {Cloud} of {Cavitation} {Bubbles}},}\ }\href {\doibase 10.1115/1.2823549} {\bibfield  {journal} {\bibinfo  {journal} {Journal of Fluids Engineering}\ }\textbf {\bibinfo {volume} {121}},\ \bibinfo {pages} {872--880} (\bibinfo {year} {1999})}\BibitemShut {NoStop}%
    \bibitem [{\citenamefont {Shen}\ \emph {et~al.}(2021)\citenamefont {Shen}, \citenamefont {Zhang}, \citenamefont {Wu},\ and\ \citenamefont {Chen}}]{Shen2021}%
      \BibitemOpen
      \bibfield  {author} {\bibinfo {author} {\bibfnamefont {Y.}~\bibnamefont {Shen}}, \bibinfo {author} {\bibfnamefont {L.}~\bibnamefont {Zhang}}, \bibinfo {author} {\bibfnamefont {Y.}~\bibnamefont {Wu}}, \ and\ \bibinfo {author} {\bibfnamefont {W.}~\bibnamefont {Chen}},\ }\bibfield  {title} {{\selectlanguage {en}\enquote {\bibinfo {title} {The role of the bubble–bubble interaction on radial pulsations of bubbles},}\ }}\href {\doibase 10.1016/j.ultsonch.2021.105535} {\bibfield  {journal} {\bibinfo  {journal} {Ultrasonics Sonochemistry}\ }\textbf {\bibinfo {volume} {73}},\ \bibinfo {pages} {105535} (\bibinfo {year} {2021})}\BibitemShut {NoStop}%
    \bibitem [{\citenamefont {Deng}\ \emph {et~al.}(2024)\citenamefont {Deng}, \citenamefont {Zhao}, \citenamefont {Zhang},\ and\ \citenamefont {Li}}]{Deng2024}%
      \BibitemOpen
      \bibfield  {author} {\bibinfo {author} {\bibfnamefont {F.}~\bibnamefont {Deng}}, \bibinfo {author} {\bibfnamefont {D.}~\bibnamefont {Zhao}}, \bibinfo {author} {\bibfnamefont {L.}~\bibnamefont {Zhang}}, \ and\ \bibinfo {author} {\bibfnamefont {Y.}~\bibnamefont {Li}},\ }\bibfield  {title} {{\selectlanguage {en}\enquote {\bibinfo {title} {Acoustic radiation of bubble clusters with different volume fractions},}\ }}\href {\doibase 10.1063/5.0195923} {\bibfield  {journal} {\bibinfo  {journal} {Physics of Fluids}\ }\textbf {\bibinfo {volume} {36}},\ \bibinfo {pages} {033308} (\bibinfo {year} {2024})}\BibitemShut {NoStop}%
    \bibitem [{\citenamefont {Rayleigh}(1917)}]{Rayleigh1917}%
      \BibitemOpen
      \bibfield  {author} {\bibinfo {author} {\bibfnamefont {L.}~\bibnamefont {Rayleigh}},\ }\bibfield  {title} {\enquote {\bibinfo {title} {On the pressure developed in a liquid during the collapse of a spherical cavity},}\ }\href {\doibase 10.1080/14786440808635681} {\bibfield  {journal} {\bibinfo  {journal} {Philosophical Magazine}\ }\textbf {\bibinfo {volume} {34}},\ \bibinfo {pages} {94--98} (\bibinfo {year} {1917})}\BibitemShut {NoStop}%
    \bibitem [{\citenamefont {Gilmore}(1952)}]{Gilmore1952}%
      \BibitemOpen
      \bibfield  {author} {\bibinfo {author} {\bibfnamefont {F.~R.}\ \bibnamefont {Gilmore}},\ }\href {https://resolver.caltech.edu/CaltechAUTHORS:Gilmore_fr_26-4} {\enquote {\bibinfo {title} {The growth or collapse of a spherical bubble in a viscous compressible liquid},}\ }\bibinfo {type} {Tech. Rep.}\ \bibinfo {number} {Report No. 26-4}\ (\bibinfo  {institution} {California Institute of Technology},\ \bibinfo {address} {Pasadena, California, USA},\ \bibinfo {year} {1952})\BibitemShut {NoStop}%
    \bibitem [{\citenamefont {Trilling}(1952)}]{Trilling1952}%
      \BibitemOpen
      \bibfield  {author} {\bibinfo {author} {\bibfnamefont {L.}~\bibnamefont {Trilling}},\ }\bibfield  {title} {{\selectlanguage {en}\enquote {\bibinfo {title} {The {Collapse} and {Rebound} of a {Gas} {Bubble}},}\ }}\href {\doibase 10.1063/1.1701962} {\bibfield  {journal} {\bibinfo  {journal} {Journal of Applied Physics}\ }\textbf {\bibinfo {volume} {23}},\ \bibinfo {pages} {14--17} (\bibinfo {year} {1952})}\BibitemShut {NoStop}%
    \bibitem [{\citenamefont {Keller}\ and\ \citenamefont {Miksis}(1980)}]{Keller1980}%
      \BibitemOpen
      \bibfield  {author} {\bibinfo {author} {\bibfnamefont {J.~B.}\ \bibnamefont {Keller}}\ and\ \bibinfo {author} {\bibfnamefont {M.}~\bibnamefont {Miksis}},\ }\bibfield  {title} {\enquote {\bibinfo {title} {Bubble oscillations of large amplitude},}\ }\href {\doibase 10.1121/1.384720} {\bibfield  {journal} {\bibinfo  {journal} {The Journal of the Acoustical Society of America}\ }\textbf {\bibinfo {volume} {68}},\ \bibinfo {pages} {628--633} (\bibinfo {year} {1980})}\BibitemShut {NoStop}%
    \bibitem [{\citenamefont {Prosperetti}\ and\ \citenamefont {Lezzi}(1986)}]{Prosperetti1986}%
      \BibitemOpen
      \bibfield  {author} {\bibinfo {author} {\bibfnamefont {A.}~\bibnamefont {Prosperetti}}\ and\ \bibinfo {author} {\bibfnamefont {A.}~\bibnamefont {Lezzi}},\ }\bibfield  {title} {{\selectlanguage {en}\enquote {\bibinfo {title} {Bubble dynamics in a compressible liquid. {Part} 1. {First}-order theory},}\ }}\href {\doibase 10.1017/S0022112086000460} {\bibfield  {journal} {\bibinfo  {journal} {Journal of Fluid Mechanics}\ }\textbf {\bibinfo {volume} {168}},\ \bibinfo {pages} {457--478} (\bibinfo {year} {1986})}\BibitemShut {NoStop}%
    \bibitem [{\citenamefont {Lezzi}\ and\ \citenamefont {Prosperetti}(1987)}]{Lezzi1987}%
      \BibitemOpen
      \bibfield  {author} {\bibinfo {author} {\bibfnamefont {A.}~\bibnamefont {Lezzi}}\ and\ \bibinfo {author} {\bibfnamefont {A.}~\bibnamefont {Prosperetti}},\ }\bibfield  {title} {{\selectlanguage {en}\enquote {\bibinfo {title} {Bubble dynamics in a compressible liquid. {Part} 2. {Second}-order theory},}\ }}\href {\doibase 10.1017/S0022112087003185} {\bibfield  {journal} {\bibinfo  {journal} {Journal of Fluid Mechanics}\ }\textbf {\bibinfo {volume} {185}},\ \bibinfo {pages} {289--321} (\bibinfo {year} {1987})}\BibitemShut {NoStop}%
    \bibitem [{\citenamefont {Denner}(2021)}]{Denner2021}%
      \BibitemOpen
      \bibfield  {author} {\bibinfo {author} {\bibfnamefont {F.}~\bibnamefont {Denner}},\ }\bibfield  {title} {{\selectlanguage {en}\enquote {\bibinfo {title} {The {Gilmore}-{NASG} model to predict single-bubble cavitation in compressible liquids},}\ }}\href {\doibase 10.1016/j.ultsonch.2020.105307} {\bibfield  {journal} {\bibinfo  {journal} {Ultrasonics Sonochemistry}\ }\textbf {\bibinfo {volume} {70}},\ \bibinfo {pages} {105307} (\bibinfo {year} {2021})}\BibitemShut {NoStop}%
    \bibitem [{\citenamefont {Fuster}, \citenamefont {Conoir},\ and\ \citenamefont {Colonius}(2014)}]{Fuster2014}%
      \BibitemOpen
      \bibfield  {author} {\bibinfo {author} {\bibfnamefont {D.}~\bibnamefont {Fuster}}, \bibinfo {author} {\bibfnamefont {J.~M.}\ \bibnamefont {Conoir}}, \ and\ \bibinfo {author} {\bibfnamefont {T.}~\bibnamefont {Colonius}},\ }\bibfield  {title} {\enquote {\bibinfo {title} {Effect of direct bubble-bubble interactions on linear-wave propagation in bubbly liquids},}\ }\href {\doibase 10.1103/PhysRevE.90.063010} {\bibfield  {journal} {\bibinfo  {journal} {Physical Review E}\ }\textbf {\bibinfo {volume} {90}},\ \bibinfo {pages} {063010} (\bibinfo {year} {2014})}\BibitemShut {NoStop}%
    \bibitem [{\citenamefont {Fuster}(2019)}]{Fuster2019}%
      \BibitemOpen
      \bibfield  {author} {\bibinfo {author} {\bibfnamefont {D.}~\bibnamefont {Fuster}},\ }\bibfield  {title} {{\selectlanguage {en}\enquote {\bibinfo {title} {A {Review} of {Models} for {Bubble} {Clusters} in {Cavitating} {Flows}},}\ }}\href {\doibase 10.1007/s10494-018-9993-4} {\bibfield  {journal} {\bibinfo  {journal} {Flow, Turbulence and Combustion}\ }\textbf {\bibinfo {volume} {102}},\ \bibinfo {pages} {497--536} (\bibinfo {year} {2019})}\BibitemShut {NoStop}%
    \bibitem [{\citenamefont {Plesset}\ and\ \citenamefont {Prosperetti}(1977)}]{Plesset1977}%
      \BibitemOpen
      \bibfield  {author} {\bibinfo {author} {\bibfnamefont {M.~S.}\ \bibnamefont {Plesset}}\ and\ \bibinfo {author} {\bibfnamefont {A.}~\bibnamefont {Prosperetti}},\ }\bibfield  {title} {\enquote {\bibinfo {title} {Bubble {Dynamics} and {Cavitation}},}\ }\href {\doibase 10.1146/annurev.fl.09.010177.001045} {\bibfield  {journal} {\bibinfo  {journal} {Annual Review of Fluid Mechanics}\ }\textbf {\bibinfo {volume} {9}},\ \bibinfo {pages} {145--185} (\bibinfo {year} {1977})}\BibitemShut {NoStop}%
    \bibitem [{\citenamefont {Lauterborn}\ and\ \citenamefont {Kurz}(2010)}]{Lauterborn2010}%
      \BibitemOpen
      \bibfield  {author} {\bibinfo {author} {\bibfnamefont {W.}~\bibnamefont {Lauterborn}}\ and\ \bibinfo {author} {\bibfnamefont {T.}~\bibnamefont {Kurz}},\ }\bibfield  {title} {{\selectlanguage {en}\enquote {\bibinfo {title} {Physics of bubble oscillations},}\ }}\href {\doibase 10.1088/0034-4885/73/10/106501} {\bibfield  {journal} {\bibinfo  {journal} {Reports on Progress in Physics}\ }\textbf {\bibinfo {volume} {73}},\ \bibinfo {pages} {106501} (\bibinfo {year} {2010})}\BibitemShut {NoStop}%
    \bibitem [{\citenamefont {Denner}(2024)}]{Denner2024a}%
      \BibitemOpen
      \bibfield  {author} {\bibinfo {author} {\bibfnamefont {F.}~\bibnamefont {Denner}},\ }\bibfield  {title} {{\selectlanguage {en}\enquote {\bibinfo {title} {The {Kirkwood}–{Bethe} hypothesis for bubble dynamics, cavitation, and underwater explosions},}\ }}\href {\doibase 10.1063/5.0209167} {\bibfield  {journal} {\bibinfo  {journal} {Physics of Fluids}\ }\textbf {\bibinfo {volume} {36}},\ \bibinfo {pages} {051302} (\bibinfo {year} {2024})}\BibitemShut {NoStop}%
    \bibitem [{\citenamefont {Fu}\ \emph {et~al.}(2023)\citenamefont {Fu}, \citenamefont {Liang}, \citenamefont {Wang}, \citenamefont {Wang}, \citenamefont {Wang}, \citenamefont {Zhang}, \citenamefont {Wang}, \citenamefont {Vogel},\ and\ \citenamefont {Yao}}]{Fu2023}%
      \BibitemOpen
      \bibfield  {author} {\bibinfo {author} {\bibfnamefont {L.}~\bibnamefont {Fu}}, \bibinfo {author} {\bibfnamefont {X.-X.}\ \bibnamefont {Liang}}, \bibinfo {author} {\bibfnamefont {S.}~\bibnamefont {Wang}}, \bibinfo {author} {\bibfnamefont {S.}~\bibnamefont {Wang}}, \bibinfo {author} {\bibfnamefont {P.}~\bibnamefont {Wang}}, \bibinfo {author} {\bibfnamefont {Z.}~\bibnamefont {Zhang}}, \bibinfo {author} {\bibfnamefont {J.}~\bibnamefont {Wang}}, \bibinfo {author} {\bibfnamefont {A.}~\bibnamefont {Vogel}}, \ and\ \bibinfo {author} {\bibfnamefont {C.}~\bibnamefont {Yao}},\ }\bibfield  {title} {{\selectlanguage {en}\enquote {\bibinfo {title} {Laser induced spherical bubble dynamics in partially confined geometry with acoustic feedback from container walls},}\ }}\href {\doibase 10.1016/j.ultsonch.2023.106664} {\bibfield  {journal} {\bibinfo  {journal} {Ultrasonics Sonochemistry}\ }\textbf {\bibinfo {volume} {101}},\ \bibinfo {pages} {106664} (\bibinfo {year} {2023})}\BibitemShut {NoStop}%
    \bibitem [{\citenamefont {Stricker}, \citenamefont {Prosperetti},\ and\ \citenamefont {Lohse}(2011)}]{Stricker2011}%
      \BibitemOpen
      \bibfield  {author} {\bibinfo {author} {\bibfnamefont {L.}~\bibnamefont {Stricker}}, \bibinfo {author} {\bibfnamefont {A.}~\bibnamefont {Prosperetti}}, \ and\ \bibinfo {author} {\bibfnamefont {D.}~\bibnamefont {Lohse}},\ }\bibfield  {title} {{\selectlanguage {en}\enquote {\bibinfo {title} {Validation of an approximate model for the thermal behavior in acoustically driven bubbles},}\ }}\href {\doibase 10.1121/1.3626132} {\bibfield  {journal} {\bibinfo  {journal} {The Journal of the Acoustical Society of America}\ }\textbf {\bibinfo {volume} {130}},\ \bibinfo {pages} {3243--3251} (\bibinfo {year} {2011})}\BibitemShut {NoStop}%
    \bibitem [{\citenamefont {Zhou}\ and\ \citenamefont {Prosperetti}(2020)}]{Zhou2020}%
      \BibitemOpen
      \bibfield  {author} {\bibinfo {author} {\bibfnamefont {G.}~\bibnamefont {Zhou}}\ and\ \bibinfo {author} {\bibfnamefont {A.}~\bibnamefont {Prosperetti}},\ }\bibfield  {title} {{\selectlanguage {en}\enquote {\bibinfo {title} {Modelling the thermal behaviour of gas bubbles},}\ }}\href {\doibase 10.1017/jfm.2020.645} {\bibfield  {journal} {\bibinfo  {journal} {Journal of Fluid Mechanics}\ }\textbf {\bibinfo {volume} {901}},\ \bibinfo {pages} {R3} (\bibinfo {year} {2020})}\BibitemShut {NoStop}%
    \bibitem [{\citenamefont {Klapcsik}\ and\ \citenamefont {Hegedűs}(2019)}]{Klapcsik2019}%
      \BibitemOpen
      \bibfield  {author} {\bibinfo {author} {\bibfnamefont {K.}~\bibnamefont {Klapcsik}}\ and\ \bibinfo {author} {\bibfnamefont {F.}~\bibnamefont {Hegedűs}},\ }\bibfield  {title} {{\selectlanguage {en}\enquote {\bibinfo {title} {Study of non-spherical bubble oscillations under acoustic irradiation in viscous liquid},}\ }}\href {\doibase 10.1016/j.ultsonch.2019.01.031} {\bibfield  {journal} {\bibinfo  {journal} {Ultrasonics Sonochemistry}\ }\textbf {\bibinfo {volume} {54}},\ \bibinfo {pages} {256--273} (\bibinfo {year} {2019})}\BibitemShut {NoStop}%
    \bibitem [{\citenamefont {Folden}\ and\ \citenamefont {Aschmoneit}(2023)}]{Folden2023}%
      \BibitemOpen
      \bibfield  {author} {\bibinfo {author} {\bibfnamefont {T.~S.}\ \bibnamefont {Folden}}\ and\ \bibinfo {author} {\bibfnamefont {F.~J.}\ \bibnamefont {Aschmoneit}},\ }\bibfield  {title} {{\selectlanguage {en}\enquote {\bibinfo {title} {A classification and review of cavitation models with an emphasis on physical aspects of cavitation},}\ }}\href {\doibase 10.1063/5.0157926} {\bibfield  {journal} {\bibinfo  {journal} {Physics of Fluids}\ }\textbf {\bibinfo {volume} {35}},\ \bibinfo {pages} {081301} (\bibinfo {year} {2023})}\BibitemShut {NoStop}%
    \bibitem [{\citenamefont {Mnich}\ \emph {et~al.}(2024)\citenamefont {Mnich}, \citenamefont {Reuter}, \citenamefont {Denner},\ and\ \citenamefont {Ohl}}]{Mnich2024}%
      \BibitemOpen
      \bibfield  {author} {\bibinfo {author} {\bibfnamefont {D.}~\bibnamefont {Mnich}}, \bibinfo {author} {\bibfnamefont {F.}~\bibnamefont {Reuter}}, \bibinfo {author} {\bibfnamefont {F.}~\bibnamefont {Denner}}, \ and\ \bibinfo {author} {\bibfnamefont {C.-D.}\ \bibnamefont {Ohl}},\ }\bibfield  {title} {{\selectlanguage {en}\enquote {\bibinfo {title} {Single cavitation bubble dynamics in a stagnation ﬂow},}\ }}\href {\doibase 10.1017/jfm.2023.1048} {\bibfield  {journal} {\bibinfo  {journal} {Journal of Fluid Mechanics}\ }\textbf {\bibinfo {volume} {979}},\ \bibinfo {pages} {A18} (\bibinfo {year} {2024})}\BibitemShut {NoStop}%
    \bibitem [{\citenamefont {Saini}\ \emph {et~al.}(2024)\citenamefont {Saini}, \citenamefont {Saade}, \citenamefont {Fuster},\ and\ \citenamefont {Lohse}}]{Saini2024}%
      \BibitemOpen
      \bibfield  {author} {\bibinfo {author} {\bibfnamefont {M.}~\bibnamefont {Saini}}, \bibinfo {author} {\bibfnamefont {Y.}~\bibnamefont {Saade}}, \bibinfo {author} {\bibfnamefont {D.}~\bibnamefont {Fuster}}, \ and\ \bibinfo {author} {\bibfnamefont {D.}~\bibnamefont {Lohse}},\ }\bibfield  {title} {{\selectlanguage {en}\enquote {\bibinfo {title} {Finite speed of sound effects on asymmetry in multibubble cavitation},}\ }}\href {\doibase 10.1103/PhysRevFluids.9.043602} {\bibfield  {journal} {\bibinfo  {journal} {Physical Review Fluids}\ }\textbf {\bibinfo {volume} {9}},\ \bibinfo {pages} {043602} (\bibinfo {year} {2024})}\BibitemShut {NoStop}%
    \bibitem [{\citenamefont {Fan}, \citenamefont {Li},\ and\ \citenamefont {Fuster}(2021)}]{Fan2021}%
      \BibitemOpen
      \bibfield  {author} {\bibinfo {author} {\bibfnamefont {Y.}~\bibnamefont {Fan}}, \bibinfo {author} {\bibfnamefont {H.}~\bibnamefont {Li}}, \ and\ \bibinfo {author} {\bibfnamefont {D.}~\bibnamefont {Fuster}},\ }\bibfield  {title} {{\selectlanguage {en}\enquote {\bibinfo {title} {Time-delayed interactions on acoustically driven bubbly screens},}\ }}\href {\doibase 10.1121/10.0008905} {\bibfield  {journal} {\bibinfo  {journal} {The Journal of the Acoustical Society of America}\ }\textbf {\bibinfo {volume} {150}},\ \bibinfo {pages} {4219--4231} (\bibinfo {year} {2021})}\BibitemShut {NoStop}%
    \bibitem [{\citenamefont {Haghi}\ and\ \citenamefont {Kolios}(2022)}]{Haghi2022}%
      \BibitemOpen
      \bibfield  {author} {\bibinfo {author} {\bibfnamefont {H.}~\bibnamefont {Haghi}}\ and\ \bibinfo {author} {\bibfnamefont {M.~C.}\ \bibnamefont {Kolios}},\ }\bibfield  {title} {{\selectlanguage {en}\enquote {\bibinfo {title} {The role of primary and secondary delays in the effective resonance frequency of acoustically interacting microbubbles},}\ }}\href {\doibase 10.1016/j.ultsonch.2022.106033} {\bibfield  {journal} {\bibinfo  {journal} {Ultrasonics Sonochemistry}\ }\textbf {\bibinfo {volume} {86}},\ \bibinfo {pages} {106033} (\bibinfo {year} {2022})}\BibitemShut {NoStop}%
    \bibitem [{\citenamefont {Mettin}\ \emph {et~al.}(1997)\citenamefont {Mettin}, \citenamefont {Akhatov}, \citenamefont {Parlitz}, \citenamefont {Ohl},\ and\ \citenamefont {Lauterborn}}]{Mettin1997}%
      \BibitemOpen
      \bibfield  {author} {\bibinfo {author} {\bibfnamefont {R.}~\bibnamefont {Mettin}}, \bibinfo {author} {\bibfnamefont {I.}~\bibnamefont {Akhatov}}, \bibinfo {author} {\bibfnamefont {U.}~\bibnamefont {Parlitz}}, \bibinfo {author} {\bibfnamefont {C.-D.}\ \bibnamefont {Ohl}}, \ and\ \bibinfo {author} {\bibfnamefont {W.}~\bibnamefont {Lauterborn}},\ }\bibfield  {title} {{\selectlanguage {en}\enquote {\bibinfo {title} {Bjerknes forces between small cavitation bubbles in a strong acoustic field},}\ }}\href {\doibase 10.1103/PhysRevE.56.2924} {\bibfield  {journal} {\bibinfo  {journal} {Physical Review E}\ }\textbf {\bibinfo {volume} {56}},\ \bibinfo {pages} {2924--2931} (\bibinfo {year} {1997})}\BibitemShut {NoStop}%
    \bibitem [{\citenamefont {Jiang}\ \emph {et~al.}(2012)\citenamefont {Jiang}, \citenamefont {Liu}, \citenamefont {Chen}, \citenamefont {Wang},\ and\ \citenamefont {Chen}}]{Jiang2012}%
      \BibitemOpen
      \bibfield  {author} {\bibinfo {author} {\bibfnamefont {L.}~\bibnamefont {Jiang}}, \bibinfo {author} {\bibfnamefont {F.}~\bibnamefont {Liu}}, \bibinfo {author} {\bibfnamefont {H.}~\bibnamefont {Chen}}, \bibinfo {author} {\bibfnamefont {J.}~\bibnamefont {Wang}}, \ and\ \bibinfo {author} {\bibfnamefont {D.}~\bibnamefont {Chen}},\ }\bibfield  {title} {{\selectlanguage {en}\enquote {\bibinfo {title} {Frequency spectrum of the noise emitted by two interacting cavitation bubbles in strong acoustic fields},}\ }}\href {\doibase 10.1103/PhysRevE.85.036312} {\bibfield  {journal} {\bibinfo  {journal} {Physical Review E}\ }\textbf {\bibinfo {volume} {85}},\ \bibinfo {pages} {036312} (\bibinfo {year} {2012})}\BibitemShut {NoStop}%
    \bibitem [{\citenamefont {Jiang}\ \emph {et~al.}(2017)\citenamefont {Jiang}, \citenamefont {Ge}, \citenamefont {Liu},\ and\ \citenamefont {Chen}}]{Jiang2017}%
      \BibitemOpen
      \bibfield  {author} {\bibinfo {author} {\bibfnamefont {L.}~\bibnamefont {Jiang}}, \bibinfo {author} {\bibfnamefont {H.}~\bibnamefont {Ge}}, \bibinfo {author} {\bibfnamefont {F.}~\bibnamefont {Liu}}, \ and\ \bibinfo {author} {\bibfnamefont {D.}~\bibnamefont {Chen}},\ }\bibfield  {title} {\enquote {\bibinfo {title} {Investigations on dynamics of interacting cavitation bubbles in strong acoustic fields},}\ }\href {\doibase 10.1016/j.ultsonch.2016.05.017} {\bibfield  {journal} {\bibinfo  {journal} {Ultrasonics Sonochemistry}\ }\textbf {\bibinfo {volume} {34}},\ \bibinfo {pages} {90--97} (\bibinfo {year} {2017})}\BibitemShut {NoStop}%
    \bibitem [{\citenamefont {Haghi}, \citenamefont {Sojahrood},\ and\ \citenamefont {Kolios}(2019)}]{Haghi2019}%
      \BibitemOpen
      \bibfield  {author} {\bibinfo {author} {\bibfnamefont {H.}~\bibnamefont {Haghi}}, \bibinfo {author} {\bibfnamefont {A.}~\bibnamefont {Sojahrood}}, \ and\ \bibinfo {author} {\bibfnamefont {M.~C.}\ \bibnamefont {Kolios}},\ }\bibfield  {title} {{\selectlanguage {en}\enquote {\bibinfo {title} {Collective nonlinear behavior of interacting polydisperse microbubble clusters},}\ }}\href {\doibase 10.1016/j.ultsonch.2019.104708} {\bibfield  {journal} {\bibinfo  {journal} {Ultrasonics Sonochemistry}\ }\textbf {\bibinfo {volume} {58}},\ \bibinfo {pages} {104708} (\bibinfo {year} {2019})}\BibitemShut {NoStop}%
    \bibitem [{\citenamefont {Guédra}, \citenamefont {Cornu},\ and\ \citenamefont {Inserra}(2017)}]{Guedra2017}%
      \BibitemOpen
      \bibfield  {author} {\bibinfo {author} {\bibfnamefont {M.}~\bibnamefont {Guédra}}, \bibinfo {author} {\bibfnamefont {C.}~\bibnamefont {Cornu}}, \ and\ \bibinfo {author} {\bibfnamefont {C.}~\bibnamefont {Inserra}},\ }\bibfield  {title} {{\selectlanguage {en}\enquote {\bibinfo {title} {A derivation of the stable cavitation threshold accounting for bubble-bubble interactions},}\ }}\href {\doibase 10.1016/j.ultsonch.2017.03.010} {\bibfield  {journal} {\bibinfo  {journal} {Ultrasonics Sonochemistry}\ }\textbf {\bibinfo {volume} {38}},\ \bibinfo {pages} {168--173} (\bibinfo {year} {2017})}\BibitemShut {NoStop}%
    \bibitem [{\citenamefont {Yasui}\ \emph {et~al.}(2009)\citenamefont {Yasui}, \citenamefont {Lee}, \citenamefont {Tuziuti}, \citenamefont {Towata}, \citenamefont {Kozuka},\ and\ \citenamefont {Iida}}]{Yasui2009}%
      \BibitemOpen
      \bibfield  {author} {\bibinfo {author} {\bibfnamefont {K.}~\bibnamefont {Yasui}}, \bibinfo {author} {\bibfnamefont {J.}~\bibnamefont {Lee}}, \bibinfo {author} {\bibfnamefont {T.}~\bibnamefont {Tuziuti}}, \bibinfo {author} {\bibfnamefont {A.}~\bibnamefont {Towata}}, \bibinfo {author} {\bibfnamefont {T.}~\bibnamefont {Kozuka}}, \ and\ \bibinfo {author} {\bibfnamefont {Y.}~\bibnamefont {Iida}},\ }\bibfield  {title} {{\selectlanguage {en}\enquote {\bibinfo {title} {Influence of the bubble-bubble interaction on destruction of encapsulated microbubbles under ultrasound},}\ }}\href {\doibase 10.1121/1.3179677} {\bibfield  {journal} {\bibinfo  {journal} {The Journal of the Acoustical Society of America}\ }\textbf {\bibinfo {volume} {126}},\ \bibinfo {pages} {973--982} (\bibinfo {year} {2009})}\BibitemShut {NoStop}%
    \bibitem [{\citenamefont {D'Agostino}\ and\ \citenamefont {Brennen}(1989)}]{DAgostino1989}%
      \BibitemOpen
      \bibfield  {author} {\bibinfo {author} {\bibfnamefont {L.}~\bibnamefont {D'Agostino}}\ and\ \bibinfo {author} {\bibfnamefont {C.~E.}\ \bibnamefont {Brennen}},\ }\bibfield  {title} {{\selectlanguage {en}\enquote {\bibinfo {title} {Linearized dynamics of spherical bubble clouds},}\ }}\href {\doibase 10.1017/S0022112089000339} {\bibfield  {journal} {\bibinfo  {journal} {Journal of Fluid Mechanics}\ }\textbf {\bibinfo {volume} {199}},\ \bibinfo {pages} {155--176} (\bibinfo {year} {1989})}\BibitemShut {NoStop}%
    \bibitem [{\citenamefont {Arora}, \citenamefont {Ohl},\ and\ \citenamefont {Lohse}(2007)}]{Arora2007}%
      \BibitemOpen
      \bibfield  {author} {\bibinfo {author} {\bibfnamefont {M.}~\bibnamefont {Arora}}, \bibinfo {author} {\bibfnamefont {C.~D.}\ \bibnamefont {Ohl}}, \ and\ \bibinfo {author} {\bibfnamefont {D.}~\bibnamefont {Lohse}},\ }\bibfield  {title} {\enquote {\bibinfo {title} {Effect of nuclei concentration on cavitation cluster dynamics},}\ }\href {\doibase 10.1121/1.2722045} {\bibfield  {journal} {\bibinfo  {journal} {The Journal of the Acoustical Society of America}\ }\textbf {\bibinfo {volume} {121}},\ \bibinfo {pages} {3432--3436} (\bibinfo {year} {2007})}\BibitemShut {NoStop}%
    \bibitem [{\citenamefont {Nasibullaeva}\ and\ \citenamefont {Akhatov}(2013)}]{Nasibullaeva2013}%
      \BibitemOpen
      \bibfield  {author} {\bibinfo {author} {\bibfnamefont {E.~S.}\ \bibnamefont {Nasibullaeva}}\ and\ \bibinfo {author} {\bibfnamefont {I.~S.}\ \bibnamefont {Akhatov}},\ }\bibfield  {title} {\enquote {\bibinfo {title} {Bubble cluster dynamics in an acoustic field},}\ }\href {\doibase 10.1121/1.4802906} {\bibfield  {journal} {\bibinfo  {journal} {The Journal of the Acoustical Society of America}\ }\textbf {\bibinfo {volume} {133}},\ \bibinfo {pages} {3727--3738} (\bibinfo {year} {2013})}\BibitemShut {NoStop}%
    \bibitem [{\citenamefont {Zhao}\ \emph {et~al.}(2020)\citenamefont {Zhao}, \citenamefont {Chen}, \citenamefont {Tao}, \citenamefont {Zhang},\ and\ \citenamefont {Wu}}]{Zhao2020c}%
      \BibitemOpen
      \bibfield  {author} {\bibinfo {author} {\bibfnamefont {G.}~\bibnamefont {Zhao}}, \bibinfo {author} {\bibfnamefont {W.}~\bibnamefont {Chen}}, \bibinfo {author} {\bibfnamefont {F.}~\bibnamefont {Tao}}, \bibinfo {author} {\bibfnamefont {L.}~\bibnamefont {Zhang}}, \ and\ \bibinfo {author} {\bibfnamefont {Y.}~\bibnamefont {Wu}},\ }\bibfield  {title} {{\selectlanguage {en}\enquote {\bibinfo {title} {Dynamics of bubbles in cavitation cloud based on lattice model},}\ }}\href {\doibase 10.1063/5.0010924} {\bibfield  {journal} {\bibinfo  {journal} {Journal of Applied Physics}\ }\textbf {\bibinfo {volume} {127}},\ \bibinfo {pages} {244701} (\bibinfo {year} {2020})}\BibitemShut {NoStop}%
    \bibitem [{\citenamefont {Ohl}\ \emph {et~al.}(2024)\citenamefont {Ohl}, \citenamefont {Rosselló}, \citenamefont {Fuster},\ and\ \citenamefont {Ohl}}]{Ohl2024}%
      \BibitemOpen
      \bibfield  {author} {\bibinfo {author} {\bibfnamefont {S.-W.}\ \bibnamefont {Ohl}}, \bibinfo {author} {\bibfnamefont {J.~M.}\ \bibnamefont {Rosselló}}, \bibinfo {author} {\bibfnamefont {D.}~\bibnamefont {Fuster}}, \ and\ \bibinfo {author} {\bibfnamefont {C.-D.}\ \bibnamefont {Ohl}},\ }\bibfield  {title} {{\selectlanguage {en}\enquote {\bibinfo {title} {Finite amplitude wave propagation through bubbly fluids},}\ }}\href {\doibase 10.1016/j.ijmultiphaseflow.2024.104826} {\bibfield  {journal} {\bibinfo  {journal} {International Journal of Multiphase Flow}\ }\textbf {\bibinfo {volume} {176}},\ \bibinfo {pages} {104826} (\bibinfo {year} {2024})}\BibitemShut {NoStop}%
    \bibitem [{\citenamefont {Bremond}\ \emph {et~al.}(2006)\citenamefont {Bremond}, \citenamefont {Arora}, \citenamefont {Ohl},\ and\ \citenamefont {Lohse}}]{Bremond2006}%
      \BibitemOpen
      \bibfield  {author} {\bibinfo {author} {\bibfnamefont {N.}~\bibnamefont {Bremond}}, \bibinfo {author} {\bibfnamefont {M.}~\bibnamefont {Arora}}, \bibinfo {author} {\bibfnamefont {C.-D.}\ \bibnamefont {Ohl}}, \ and\ \bibinfo {author} {\bibfnamefont {D.}~\bibnamefont {Lohse}},\ }\bibfield  {title} {\enquote {\bibinfo {title} {Controlled {Multibubble} {Surface} {Cavitation}},}\ }\href {\doibase 10.1103/PhysRevLett.96.224501} {\bibfield  {journal} {\bibinfo  {journal} {Physical Review Letters}\ }\textbf {\bibinfo {volume} {96}},\ \bibinfo {pages} {224501} (\bibinfo {year} {2006})}\BibitemShut {NoStop}%
    \bibitem [{\citenamefont {Minnaert}(1933)}]{Minnaert1933}%
      \BibitemOpen
      \bibfield  {author} {\bibinfo {author} {\bibfnamefont {M.}~\bibnamefont {Minnaert}},\ }\bibfield  {title} {{\selectlanguage {en}\enquote {\bibinfo {title} {{XVI}. \textit{{On} musical air-bubbles and the sounds of running water}},}\ }}\href {\doibase 10.1080/14786443309462277} {\bibfield  {journal} {\bibinfo  {journal} {The London, Edinburgh, and Dublin Philosophical Magazine and Journal of Science}\ }\textbf {\bibinfo {volume} {16}},\ \bibinfo {pages} {235--248} (\bibinfo {year} {1933})}\BibitemShut {NoStop}%
    \bibitem [{\citenamefont {Prosperetti}(1984)}]{Prosperetti1984a}%
      \BibitemOpen
      \bibfield  {author} {\bibinfo {author} {\bibfnamefont {A.}~\bibnamefont {Prosperetti}},\ }\bibfield  {title} {{\selectlanguage {en}\enquote {\bibinfo {title} {Bubble phenomena in sound fields: part one},}\ }}\href {\doibase 10.1016/0041-624X(84)90024-6} {\bibfield  {journal} {\bibinfo  {journal} {Ultrasonics}\ }\textbf {\bibinfo {volume} {22}},\ \bibinfo {pages} {69--77} (\bibinfo {year} {1984})}\BibitemShut {NoStop}%
    \bibitem [{\citenamefont {Denner}\ and\ \citenamefont {Schenke}(2023{\natexlab{a}})}]{Denner2023}%
      \BibitemOpen
      \bibfield  {author} {\bibinfo {author} {\bibfnamefont {F.}~\bibnamefont {Denner}}\ and\ \bibinfo {author} {\bibfnamefont {S.}~\bibnamefont {Schenke}},\ }\bibfield  {title} {{\selectlanguage {en}\enquote {\bibinfo {title} {Modeling acoustic emissions and shock formation of cavitation bubbles},}\ }}\href {\doibase 10.1063/5.0131930} {\bibfield  {journal} {\bibinfo  {journal} {Physics of Fluids}\ }\textbf {\bibinfo {volume} {35}},\ \bibinfo {pages} {012114} (\bibinfo {year} {2023}{\natexlab{a}})}\BibitemShut {NoStop}%
    \bibitem [{\citenamefont {Fuster}\ and\ \citenamefont {Colonius}(2011)}]{Fuster2011a}%
      \BibitemOpen
      \bibfield  {author} {\bibinfo {author} {\bibfnamefont {D.}~\bibnamefont {Fuster}}\ and\ \bibinfo {author} {\bibfnamefont {T.}~\bibnamefont {Colonius}},\ }\bibfield  {title} {{\selectlanguage {en}\enquote {\bibinfo {title} {Modelling bubble clusters in compressible liquids},}\ }}\href {\doibase 10.1017/jfm.2011.380} {\bibfield  {journal} {\bibinfo  {journal} {Journal of Fluid Mechanics}\ }\textbf {\bibinfo {volume} {688}},\ \bibinfo {pages} {352--389} (\bibinfo {year} {2011})}\BibitemShut {NoStop}%
    \bibitem [{\citenamefont {Zhang}\ \emph {et~al.}(2023)\citenamefont {Zhang}, \citenamefont {Li}, \citenamefont {Cui}, \citenamefont {Li},\ and\ \citenamefont {Liu}}]{Zhang2023b}%
      \BibitemOpen
      \bibfield  {author} {\bibinfo {author} {\bibfnamefont {A.-M.}\ \bibnamefont {Zhang}}, \bibinfo {author} {\bibfnamefont {S.-M.}\ \bibnamefont {Li}}, \bibinfo {author} {\bibfnamefont {P.}~\bibnamefont {Cui}}, \bibinfo {author} {\bibfnamefont {S.}~\bibnamefont {Li}}, \ and\ \bibinfo {author} {\bibfnamefont {Y.-L.}\ \bibnamefont {Liu}},\ }\bibfield  {title} {{\selectlanguage {en}\enquote {\bibinfo {title} {A unified theory for bubble dynamics},}\ }}\href {\doibase 10.1063/5.0145415} {\bibfield  {journal} {\bibinfo  {journal} {Physics of Fluids}\ }\textbf {\bibinfo {volume} {35}},\ \bibinfo {pages} {033323} (\bibinfo {year} {2023})}\BibitemShut {NoStop}%
    \bibitem [{\citenamefont {Denner}\ and\ \citenamefont {Schenke}(2023{\natexlab{b}})}]{Denner2023a}%
      \BibitemOpen
      \bibfield  {author} {\bibinfo {author} {\bibfnamefont {F.}~\bibnamefont {Denner}}\ and\ \bibinfo {author} {\bibfnamefont {S.}~\bibnamefont {Schenke}},\ }\bibfield  {title} {{\selectlanguage {en}\enquote {\bibinfo {title} {{APECSS}: {A} software library for cavitation bubble dynamics and acoustic emissions},}\ }}\href {\doibase 10.21105/joss.05435} {\bibfield  {journal} {\bibinfo  {journal} {Journal of Open Source Software}\ }\textbf {\bibinfo {volume} {8}},\ \bibinfo {pages} {5435} (\bibinfo {year} {2023}{\natexlab{b}})}\BibitemShut {NoStop}%
    \bibitem [{\citenamefont {Maeda}\ and\ \citenamefont {Colonius}(2019)}]{Maeda2019}%
      \BibitemOpen
      \bibfield  {author} {\bibinfo {author} {\bibfnamefont {K.}~\bibnamefont {Maeda}}\ and\ \bibinfo {author} {\bibfnamefont {T.}~\bibnamefont {Colonius}},\ }\bibfield  {title} {{\selectlanguage {en}\enquote {\bibinfo {title} {Bubble cloud dynamics in an ultrasound field},}\ }}\href {\doibase 10.1017/jfm.2018.968} {\bibfield  {journal} {\bibinfo  {journal} {Journal of Fluid Mechanics}\ }\textbf {\bibinfo {volume} {862}},\ \bibinfo {pages} {1105--1134} (\bibinfo {year} {2019})}\BibitemShut {NoStop}%
    \bibitem [{\citenamefont {Maeda}\ and\ \citenamefont {Colonius}(2017)}]{Maeda2017}%
      \BibitemOpen
      \bibfield  {author} {\bibinfo {author} {\bibfnamefont {K.}~\bibnamefont {Maeda}}\ and\ \bibinfo {author} {\bibfnamefont {T.}~\bibnamefont {Colonius}},\ }\bibfield  {title} {\enquote {\bibinfo {title} {A source term approach for generation of one-way acoustic waves in the {Euler} and {Navier}–{Stokes} equations},}\ }\href {\doibase 10.1016/j.wavemoti.2017.08.004} {\bibfield  {journal} {\bibinfo  {journal} {Wave Motion}\ }\textbf {\bibinfo {volume} {75}},\ \bibinfo {pages} {36--49} (\bibinfo {year} {2017})}\BibitemShut {NoStop}%
    \bibitem [{\citenamefont {Fuster}, \citenamefont {Pham},\ and\ \citenamefont {Zaleski}(2014)}]{Fuster2014a}%
      \BibitemOpen
      \bibfield  {author} {\bibinfo {author} {\bibfnamefont {D.}~\bibnamefont {Fuster}}, \bibinfo {author} {\bibfnamefont {K.}~\bibnamefont {Pham}}, \ and\ \bibinfo {author} {\bibfnamefont {S.}~\bibnamefont {Zaleski}},\ }\bibfield  {title} {{\selectlanguage {en}\enquote {\bibinfo {title} {Stability of bubbly liquids and its connection to the process of cavitation inception},}\ }}\href {\doibase 10.1063/1.4870820} {\bibfield  {journal} {\bibinfo  {journal} {Physics of Fluids}\ }\textbf {\bibinfo {volume} {26}},\ \bibinfo {pages} {042002} (\bibinfo {year} {2014})}\BibitemShut {NoStop}%
    \bibitem [{\citenamefont {Neppiras}\ and\ \citenamefont {Noltingk}(1951)}]{Neppiras1951}%
      \BibitemOpen
      \bibfield  {author} {\bibinfo {author} {\bibfnamefont {E.~A.}\ \bibnamefont {Neppiras}}\ and\ \bibinfo {author} {\bibfnamefont {B.~E.}\ \bibnamefont {Noltingk}},\ }\bibfield  {title} {{\selectlanguage {en}\enquote {\bibinfo {title} {Cavitation {Produced} by {Ultrasonics}: {Theoretical} {Conditions} for the {Onset} of {Cavitation}},}\ }}\href {\doibase 10.1088/0370-1301/64/12/302} {\bibfield  {journal} {\bibinfo  {journal} {Proceedings of the Physical Society. Section B}\ }\textbf {\bibinfo {volume} {64}},\ \bibinfo {pages} {1032--1038} (\bibinfo {year} {1951})}\BibitemShut {NoStop}%
    \bibitem [{\citenamefont {Ando}, \citenamefont {Colonius},\ and\ \citenamefont {Brennen}(2011)}]{Ando2011}%
      \BibitemOpen
      \bibfield  {author} {\bibinfo {author} {\bibfnamefont {K.}~\bibnamefont {Ando}}, \bibinfo {author} {\bibfnamefont {T.}~\bibnamefont {Colonius}}, \ and\ \bibinfo {author} {\bibfnamefont {C.~E.}\ \bibnamefont {Brennen}},\ }\bibfield  {title} {\enquote {\bibinfo {title} {Numerical simulation of shock propagation in a polydisperse bubbly liquid},}\ }\href {\doibase 10.1016/j.ijmultiphaseflow.2011.03.007} {\bibfield  {journal} {\bibinfo  {journal} {International Journal of Multiphase Flow}\ }\textbf {\bibinfo {volume} {37}},\ \bibinfo {pages} {596--608} (\bibinfo {year} {2011})}\BibitemShut {NoStop}%
    \bibitem [{\citenamefont {Wagner}\ and\ \citenamefont {Pruß}(2002)}]{Wagner2002}%
      \BibitemOpen
      \bibfield  {author} {\bibinfo {author} {\bibfnamefont {W.}~\bibnamefont {Wagner}}\ and\ \bibinfo {author} {\bibfnamefont {A.}~\bibnamefont {Pruß}},\ }\bibfield  {title} {\enquote {\bibinfo {title} {The {IAPWS} {Formulation} 1995 for the {Thermodynamic} {Properties} of {Ordinary} {Water} {Substance} for {General} and {Scientific} {Use}},}\ }\href {\doibase 10.1063/1.1461829} {\bibfield  {journal} {\bibinfo  {journal} {Journal of Physical and Chemical Reference Data}\ }\textbf {\bibinfo {volume} {31}},\ \bibinfo {pages} {387--535} (\bibinfo {year} {2002})}\BibitemShut {NoStop}%
    \end{thebibliography}

%
    
\end{document}